\documentclass[reprint,
 superscriptaddress,
 groupedaddress,
preprintnumbers,
nofootinbib,
floatfix,
longbibliography,
 amsmath,amssymb,
 aps,
 showpacs,
 prd,
floatfix,
]{revtex4-2}

\usepackage{booktabs}

\usepackage{xspace}
\usepackage{acronym}

\usepackage{color, colortbl}
\definecolor{Gray}{gray}{0.9}
\usepackage{siunitx}
\usepackage{makecell}
\usepackage{multirow}
\usepackage{float}

\usepackage{hyperref}
\hypersetup{
  colorlinks   = true,
  urlcolor     = blue,
  linkcolor    = blue,
  citecolor    = red
}
\usepackage{graphicx}
\usepackage{dcolumn}
\usepackage{bm}
\usepackage{caption}
\usepackage{subcaption}
\usepackage[normalem]{ulem}

\usepackage{xcolor}
\usepackage{orcidlink}

\begin{document}

\preprint{APS/123-QED}
\title{Detectability of eccentric binary black holes with PyCBC and cWB pipelines during the third observing run of LIGO-Virgo-KAGRA}

\author{Bhooshan Gadre\orcidlink{0000-0002-1534-9761}}
\email{b.u.gadre@uu.nl}
 \affiliation{Institute for Gravitational and Subatomic Physics (GRASP),
Utrecht University, Princetonplein 1, 3584 CC Utrecht, The Netherlands}

\author{Kanchan Soni\orcidlink{0000-0001-8051-7883}}
\affiliation{Department of Physics, Syracuse University, Syracuse, NY 13244, USA}
\affiliation{Inter-University Centre for Astronomy and Astrophysics, Pune 411007, India}

\author{Shubhanshu Tiwari\orcidlink{0000-0003-1611-6625}}
\affiliation{%
 Physik-Institut, Universit\"{a}t Z\"{u}rich, Winterthurerstrasse 190, 8057 Z\"{u}rich, Switzerland
}%

\author{Antoni Ramos-Buades\orcidlink{0000-0002-6874-7421}}
\affiliation{Nikhef, Science Park 105, 1098 XG, Amsterdam, The Netherlands}
\author{Maria Haney\orcidlink{0000-0001-7554-3665}}
\affiliation{Nikhef, Science Park 105, 1098 XG, Amsterdam, The Netherlands}

\author{Sanjit Mitra}
\affiliation{
Inter-University Centre for Astronomy and Astrophysics, Pune 411007, India}

\begin{abstract}

Detecting binary black hole (BBH) mergers with quantifiable orbital eccentricity would confirm the existence of a dynamical formation channel for these binaries. The current state-of-the-art gravitational wave searches of LIGO-Virgo-KAGRA strain data focus more on quasicircular mergers due to increased dimensionality and lack of efficient eccentric waveform models.  In this work, we compare the sensitivities of two search pipelines, the matched filter-based \texttt{PyCBC} and the unmodelled coherent Wave Burst (\texttt{cWB}) algorithms towards the spinning eccentric BBH mergers, using a multipolar nonprecessing-spin eccentric signal model, \texttt{SEOBNRv4EHM}. Our findings show that neglecting eccentricity leads to missed opportunities for detecting eccentric BBH mergers, with \texttt{PyCBC} exhibiting a $10-20\, \%$ sensitivity loss for eccentricities exceeding $0.2$ defined at $10$ Hz. In contrast, \texttt{cWB} is resilient, with a $10\, \%$ sensitivity increase for heavier ($\mathcal{M} \ge 30 \, \text{M}_{\odot}$) eccentric BBH mergers, but is significantly less sensitive than \texttt{PyCBC} for lighter BBH mergers. Our fitting factor study confirmed that neglecting eccentricity biases the estimation of chirp mass, mass ratio, and effective spin parameter, skewing our understanding of astrophysical BBH populations, fundamental physics, and precision cosmology. Our results demonstrate that the current search pipelines are not sufficiently sensitive to eccentric BBH mergers, necessitating the development of a dedicated matched-filter search for these binaries. Whereas, burst searches should be optimized to detect lower chirp mass BBH mergers as eccentricity does not affect their search sensitivity significantly.

\end{abstract}

\maketitle

\acrodef{LVK}[LVK]{LIGO-Virgo-KAGRA Collaboration}
\acrodef{BH}[BH]{black hole}
\acrodef{BBH}[BBH]{binary black hole}
\acrodef{CBC}[CBC]{compact binary coalescence}
\acrodef{GW}[GW]{gravitational wave}
\acrodef{CWB}[cWB]{coherent WaveBurst}
\acrodef{PSO}[PSO]{\emph{Particle Swarm Optimisation}\xspace}
\acrodef{DiffEvol}[DiffEvol]{\emph{Differential Evolution}\xspace}
\acrodef{NR}[NR]{numerical relativity}
\acrodef{FAR}[FAR]{false alarm rate}
\acrodef{AGN}[AGN]{Active Galactic Nuclei}
\acrodef{aLIGO}[aLIGO]{Advanced Laser Interferometer Gravitational wave Observatory}
\acrodef{Virgo}[Virgo]{Advanced Virgo}
\acrodef{IMR}[IMR]{inspiral-merger-ringdown}
\acrodef{BNS}[BNS]{binary neutron star}
\acrodef{NS}[NS]{neutron star}

\newcommand{\gw}[0]{\ac{GW}\xspace}
\newcommand{\lvk}[0]{\ac{LVK}\xspace}
\newcommand{\msun}[0]{\text{M}_{\odot}\xspace}
\newcommand{\pycbc}[0]{\texttt{PyCBC}\xspace}
\newcommand{\cwb}[0]{\texttt{cWB}\xspace}
\newcommand{\bbh}[0]{\ac{BBH}\xspace}
\newcommand{\bh}[0]{\ac{BH}\xspace}
\newcommand{\cbc}[0]{\ac{CBC}\xspace}
\newcommand{\seob}[0]{\texttt{SEOBNRv4E}\xspace}
\newcommand{\seobhm}[0]{\texttt{SEOBNRv4EHM}\xspace}
\newcommand{\ff}{$\mathcal{FF}$\xspace}
\newcommand{\pso}[0]{\texttt{PSO}\xspace}
\newcommand{\diffevol}[0]{\texttt{DiffEvol}\xspace}
\newcommand{\NR}[0]{\ac{NR}\xspace}
\newcommand{\far}[0]{\ac{FAR}\xspace}
\newcommand{\agn}[0]{\ac{AGN}\xspace}
\newcommand{\aLIGO}[0]{\ac{aLIGO}\xspace}
\newcommand{\Virgo}[0]{\ac{Virgo}\xspace}
\newcommand{\IMR}[0]{\ac{IMR}\xspace}

\section{\label{sec:intro}Introduction}

The field of gravitational-wave science boomed with the first detection~\cite{gw150914} of these then-elusive signals. Since then, the \lvk has successfully detected more than 90 GW signals~\cite{gwtc3_paper}, predominantly originating from the mergers of \bbh. Additionally, several independent groups~\cite{nitz_ogc3,venumadhav_2022_o3a} have catalogued these events, including some below the detection threshold, by utilizing publicly available data~\cite{LIGOScientific:2019lzm}. Despite the substantial number of detections, our understanding of the astrophysical origins of these signals remains incomplete.

\begin{figure*}[htb]
    \centering
    \includegraphics[width=\textwidth]{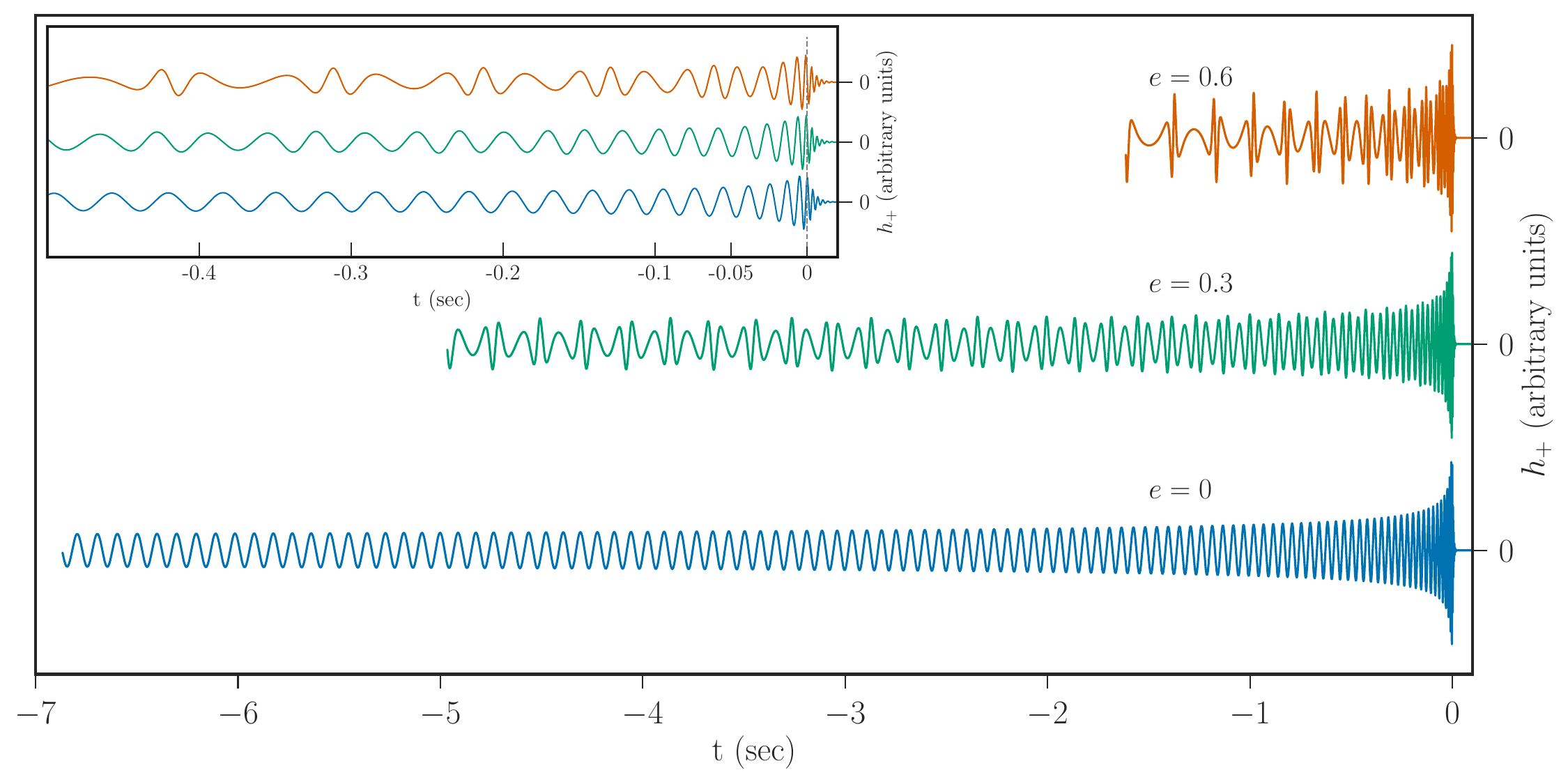}
    \caption{\texttt{SEOBNRv4EHM} waveforms for merging BBH with the same initial conditions except for eccentricities. The inset shows the zoomed-in version of the same waveforms. The fixed parameters used in the waveform generation are $\left(m_1\,, m_2\,, s_{1z}\,, s_{2z}\,, \iota\,, \phi_{{\rm ref}}\,, f_{{\rm ref}} \right) = \left(40 \, \msun\,, 20 \, \msun\,, 0.3\,, 0.1\,, 2\pi/3\,, 0\,, 10 ~{\rm Hz}\right)$.}
    \label{fig:waveforms}
\end{figure*}

There are a few possibilities for the formation of a binary black hole. It can either occur through the evolution of isolated star binaries~\cite{bethe_1998} or as a result of dynamical interactions in stellar clusters or triple systems~\cite{Barack_2019}. Another formation mechanism involves gas capture in \agn-assisted mergers~\cite{McKernan_2012}. While isolated \bbh can initially possess significant eccentricity, the emission of \gw gradually circularizes their orbits as their orbital frequency aligns with ground-based gravitational wave observatories~\cite{peters_two_point_gw}. Conversely, dynamically formed binaries resulting from close encounters tend to approach mergers more closely, leaving limited time for orbital circularization. Notably, within \agn disks, the eccentricity of mergers can be significantly enhanced through interactions with a nearby third object, primarily facilitated by the Kozai-Lidov mechanism~\cite{wen2003_KL_mechanism}. Therefore, the characterization of orbital eccentricity ($e$), or its absence, within populations of BBHs yields critical insights into the relative contributions of various formation channels for these binaries~\cite{Zevin:2021rtf,Zeeshan:2024ovp}.

According to many studies, none of the detected \bbh mergers has exhibited clear evidence of the presence of eccentricity~\cite{OShea:2021faf,Romero-Shaw:2020thy,Romero-Shaw:2021ual,Gayathri:2020coq,Gamba:2021gap, Iglesias:2022xfc,Ramos-Buades:2023yhy,Wu:2020zwr}. This can be attributed to the circularizing effect of GW emission on the orbit of a binary system, particularly when the signal enters the sensitive frequency bands of detectors like Advanced LIGO~\cite{asi} and Advanced Virgo~\cite{advancedvirgo}.
However, there is a recent study claiming evidence of eccentricity in \bbh mergers observed by \lvk~\cite{Gupte:2024jfe}. There are specific scenarios where deviations from quasicircular orbits can occur. This effect can happen in environments such as globular clusters~\cite{OLeary_2006,PhysRevLett.121.161103,PhysRevD.97.103014}, where the \bbh can acquire triple companions or their orbits can get perturbed through dynamic encounters. These interactions can induce eccentricities in the lower frequency range that can align with the sensitivity limits of current detectors; approximately $10\, \%$ of sources exhibiting a moderate eccentricity of $e=0.1$ at 10 Hz~\cite{MOCCA_SURVEY_gc_bbh_ecc}.  Another possible scenario is the Kozai–Lidov oscillation mechanism of hierarchical triple systems~\cite{wen2003_KL_mechanism}, where a distant third object perturbs a binary, which can lead to high eccentricities approaching unity. In addition, hierarchical triple configurations can also form when \bbh orbits supermassive black holes in galactic nuclei. The merger of heavy stellar-mass \bh and intermediate-mass black holes is expected to occur at eccentricities greater than 0.1 measured at 10 Hz. On the other hand, low-mass binaries are anticipated to exhibit eccentricities on the order of $10^{-3}$~\cite{laszlo_ecc_upperlimit} within the LIGO-Virgo observing band.

There are two main approaches to search for \bbh mergers. The first one involves correlating the modeled GW signal with the strain data recorded by the interferometer, known as matched filtering~\cite{matched_filtering_sathya_svd,svd-sathya_1994,svd-schutz_1994}. The second approach looks for coherent excess power across the detector network~\cite{klimenko2016method,drago2021coherent}. Particularly, the Coherent Wave Burst (\cwb) algorithm, as implemented in the \cwb pipeline~\cite{klimenko2016method}, has been used for eccentric \bbh searches~\cite{Abbott_2019_ebbh,lvk_ebbh_2023}. However, no dedicated matched filter-based search has been conducted for these sources. This is significant given that matched filter searches have been shown to be more efficient than \cwb searches for \bbh mergers involving component masses less than $40\, \msun$~\cite{gwtc3_paper}. 
The reason is searching for eccentric mergers using a dedicated matched filter-based search pipeline requires the construction of an eccentric template bank. One major obstacle in generating it is the lack of accurate and fast eccentric waveform models for matched filtering. In recent years, there has been a lot of progress in producing inspiral~\cite{Moore:2016qxz,Tanay:2016zog,Moore:2019xkm,Tiwari:2019jtz,Ebersold:2019kdc,Cho:2021oai} and inspiral-merger-ringdown (IMR)~\cite{Huerta:2017kez,Hinder:2017sxy,Cao:2017ndf,Nagar:2021gss, Setyawati:2021gom, Islam:2021mha, Ramos-Buades:2021adz, Liu:2023ldr} eccentric waveforms models. Although accurate against eccentric numerical relativity (NR) waveforms, the latter has not yet been shown to be computationally efficient enough to be used to construct eccentric template banks.
Another challenge arises from the extension of the search parameter space. One must consider eccentricity and mean anomaly parameters to identify an eccentric orbit. The mean anomaly represents the fraction of the orbital period that has passed since the last closest approach in the orbit. Including these two parameters in the search space, alongside the masses and spins of the compact objects, contributes to an increase in the dimensionality and the number of templates within the search bank. Consequently, the matched filtering-based searches become more computationally intensive.

There are attempts to include eccentricity in matched filter-based search pipelines like \pycbc~\cite{Usman:2015kfa,pycbc_gareth_3det} utilizing data from the current generation of gravitational wave detectors for inspiraling subsolar mass \bh~\cite{Nitz:2022ltl} and binary neutron star~\cite{Nitz_2020_ebns} and neutron star-black hole mergers~\cite{Dhurkunde:2023qoe}, using the inspiral-only waveform model \texttt{TaylorF2e}~\cite{Moore:2016qxz}, without and with moderate spins of the binaries respectively. Both of the aforementioned \pycbc searches employed the template bank-based coincident search approach. Recently, a swarm-intelligent algorithm~\cite{Pal:2023dyg} was used to search for eccentric \bbh mergers ($e \le 0.5$) in the total mass range of $10-100 \, \msun$ and moderate non-precessing spins ($-0.5 < S_{1,2z} < 0.5$), using the IMR eccentric \texttt{TEOBResumS-DALI} waveform model~\cite{Nagar:2021gss} and \pycbc. The swarm-intelligent search is non-bank based, and hence, it is computationally less expensive. However, the cost reduction comes at the expense of the robust significance estimation.

In addition, there is a recent sensitivity study using NR waveforms as injections quantifying the effect of missing eccentricities using \pycbc and \cwb searches~\cite{Ramos-Buades:2020eju}. However, the study does not systematically cover the complete \bbh merger parameter space due to the limited availability of the eccentric NR waveforms. Hence, in the work presented here, we compare the sensitivities of \pycbc and \cwb searches to the mock eccentric signal generated with the multipolar nonprecessing-spin eccentric waveform model, \seobhm~\cite{Khalil:2021txt,Ramos-Buades:2021adz,Ramos-Buades:2023yhy}. The respective configurations of both pipelines are the same as those deployed by the LVK collaboration while hunting for GWs during the third observing run (O3).

We organize the paper as follows: In Sec.~\ref{subsec:eccWaveforms}, we briefly discuss the \seobhm waveform model, and in Sec.~\ref{subsec:injections}, we give details of our injection distribution and discuss findings of our fitting factor study. In subsequent sections ~\ref{subsec:data},~\ref{subsec:pycbc_config} and ~\ref{subsec:cwb_config}, we give details of the O3 data we have used in the study as well as configuration details of each of the pipelines. We discuss our search results for individual pipelines and comparison in Sec.\ref{sec:results}. We devote Sec.~\ref{sec:conclusion}
to summarize our conclusions.

\section{\label{sec:Method}Method}

\subsection{Gravitational waves from eccentric binaries} \label{subsec:eccWaveforms}

Gravitational waves from binaries in elliptical orbits can be described by 17 parameters. These parameters are typically divided into extrinsic and intrinsic parameters.

For generic binaries the intrinsic parameters describing the source frame are the component masses $m_{i}$, dimensionless spin vectors $\bm \chi_{i}=\vec{S}_{i}/m^2_{i}$, where $\vec{S}_i$ is the spin vector and $i=1,2$, the orbital eccentricity $e$ and a radial phase parameter. In this work, we restrict to align-spin binaries, which reduces the parameter space from 9 to 5 intrinsic parameters as the only non-zero component of the dimensionless spin vectors are the ones aligned with the orbital angular momentum of the system, i.e., $\chi_i \equiv \chi_{z,i}$.

The extrinsic parameters relating the source and the detector frames are the angular position of the line of sight measured in the source frame, given by the inclination and azimuthal angles $(\iota, \varphi)$, the sky location of the source in the detector frame $(\theta, \phi)$, the polarization angle $\psi$, the luminosity distance $d_L$ and the coalescence time $t_c$.

To produce mock eccentric GW signals, we employ the eccentric IMR waveform model \texttt{SEOBNRv4EHM}~\cite{Khalil:2021txt,Ramos-Buades:2021adz,Ramos-Buades:2023yhy}, which describes elliptical orbits using two eccentric parameters: the initial orbital eccentricity and the relativistic anomaly $\zeta$. In Fig~\ref{fig:waveforms}, we show $h_{+}$ polarisation eccentric non-precessing \bbh mergers with varying eccentricity while keeping all the other parameters fixed. With increasing eccentricity, waveforms become shorter even though they radiate more energy. The inset shows the zoomed-in version of the same waveforms. The bursts of emission due to periastron passages are visible for $e_{10{\rm Hz}} = 0.6$ (orange curve). The  \texttt{SEOBNRv4EHM} is built upon the accurate multipolar quasicircular \texttt{SEOBNRv4HM} model~\cite{Cotesta:2018fcv} with non-precessing spins, and it includes eccentric corrections up to second post-Newtonian (2PN) order~\cite{Khalil:2021txt} in the $(l,|m|)=(2,2),(2,1),(3,3),(4,4),(5,5)$ multipoles. When restricting to the $(l,|m|)=(2,2)$ modes, we refer to the model as \texttt{SEOBNRv4E}.

The  \texttt{SEOBNRv4EHM} includes eccentric effects in the inspiral effective-one-body (EOB) multipoles and employs the same merger-ringdown model as the quasicircular \texttt{SEOBNRv4HM}. Thus, the model assumes that the effects of eccentricity during merger and ringdown are negligible and that the binary has circularized by the time of the coalescence. In Ref.~\cite{Ramos-Buades:2021adz}, the \texttt{SEOBNRv4EHM} model was shown to accurately recover the quasicircular limit of the \texttt{SEOBNRv4HM} model and be accurate with an unfaithfulness $<1\, \%$ against a dataset of eccentric \NR waveforms from the Simulating eXtreme Spacetimes (SXS) catalogue~\cite{Boyle:2019kee} with moderate initial eccentricities $e_0 \leq 0.3$. Recently, \texttt{SEOBNRv4EHM} was also successfully employed in Bayesian inference studies to measure eccentricity from GW events reported by the LVK collaboration~\cite{Ramos-Buades:2023yhy}, demonstrating the ability and robustness of the model for its application in data analysis.

In this work, we use the eccentricity and relativistic anomaly definition based on the initial conditions of \seobhm~\cite{Ramos-Buades:2021adz,Ramos-Buades:2023yhy}. As eccentricity is not uniquely defined in general relativity, several definitions exist in the literature (see Ref.~\cite{Loutrel:2018ydu} for a brief summary). Recent work in the literature has focused on adopting a common definition of eccentricity measured from the GW signal~\cite{Ramos-Buades:2022lgf,Shaikh:2023ypz} with a correct Newtonian limit of eccentricity. We leave for future work adopting such a definition of eccentricity to perform sensitivity studies of search algorithms.

\subsection{Injection set}
\label{subsec:injections}

\begin{figure}[htb]
    \centering
    \includegraphics[scale=.29]{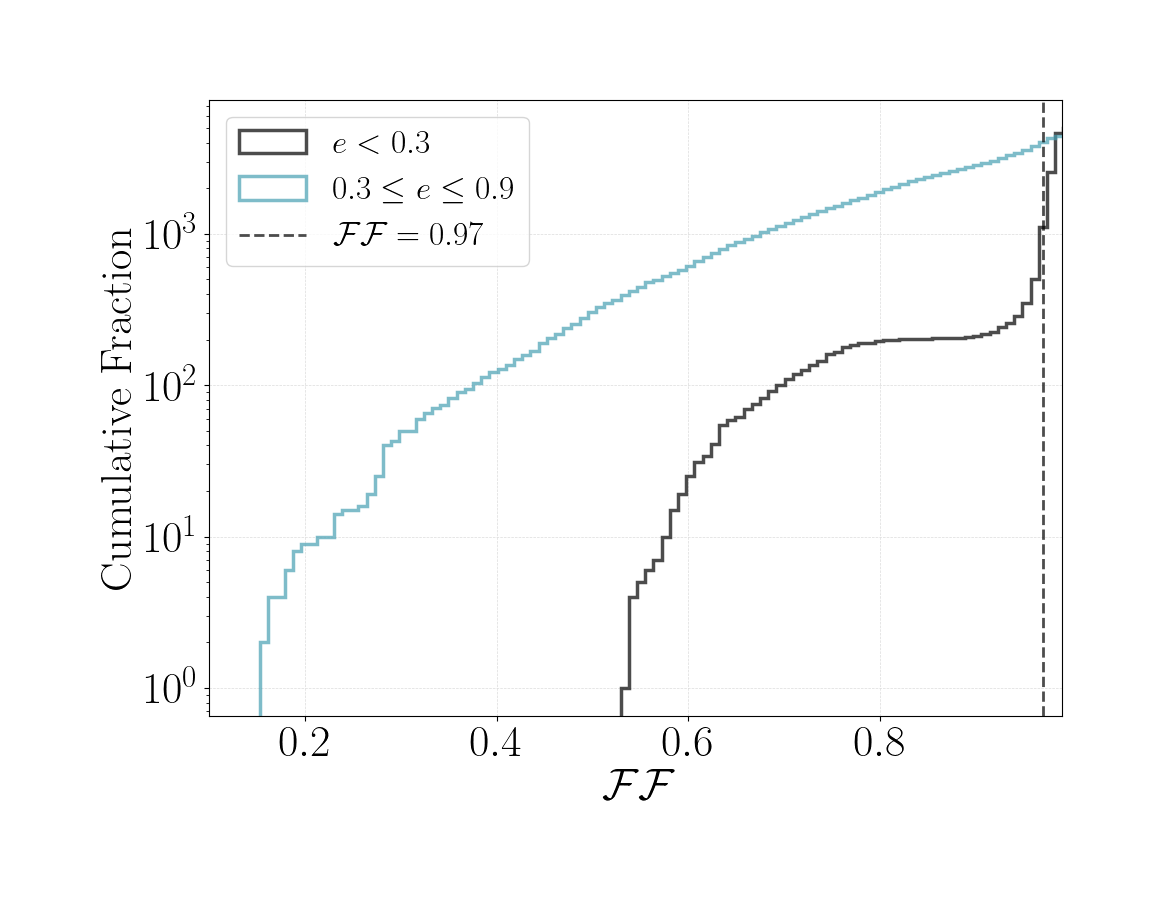}
    \caption{Effectualness of broad-\pycbc template bank computed for a simulated population of eccentric binary sources. The two distributions show the \ff for low (in black) and high (in blue) eccentricity range. The vertical line shows the lower limit for an effectual template bank used for the broad-\pycbc analysis in GWTC-3~\cite{gwtc3_paper}.}
    \label{fig:cumulative_plot_inj1inj2_comb}
\end{figure}

\begin{figure*}[htb]
    \centering
    \includegraphics[width=\textwidth]{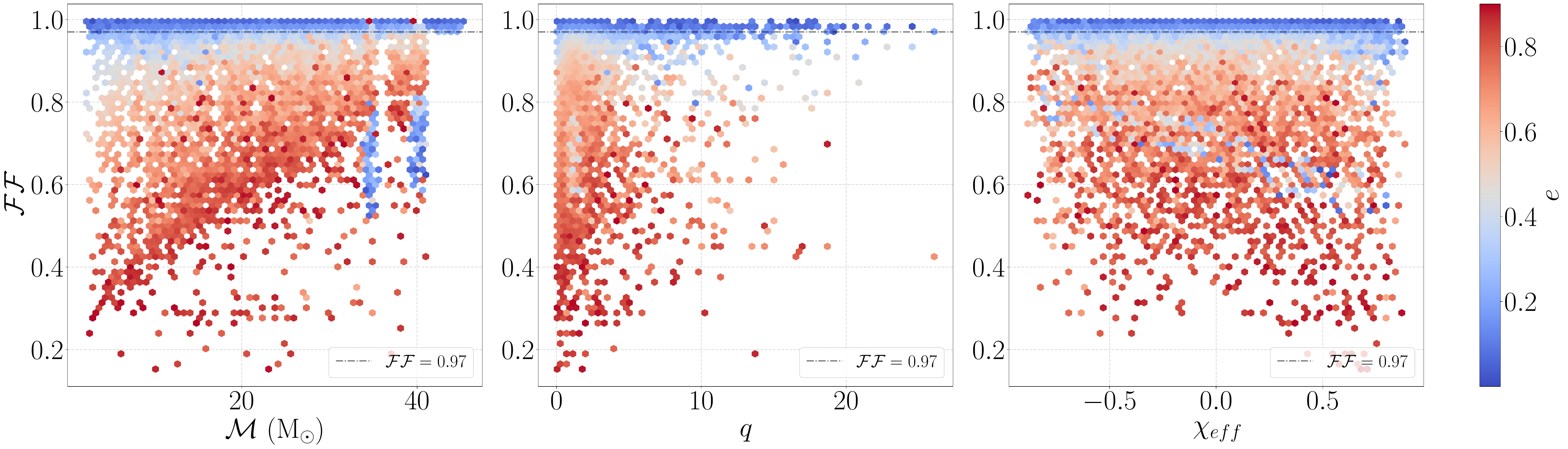}
    \caption{Recovered \ff values as a function of chirp mass ($\mathcal{M}$), mass ratio ($q$), and effective spin ($\chi_{eff}$) computed for \seob injections against \texttt{SEOBNRv4\_ROM} using template bank, The color bar indicates the range of eccentricities. A dotted horizontal line indicates the lower limit for an effectual template bank.}
    \label{fig:ff_bank}
\end{figure*}

\begin{figure*}[htb]
    \centering
    \includegraphics[width=\textwidth]{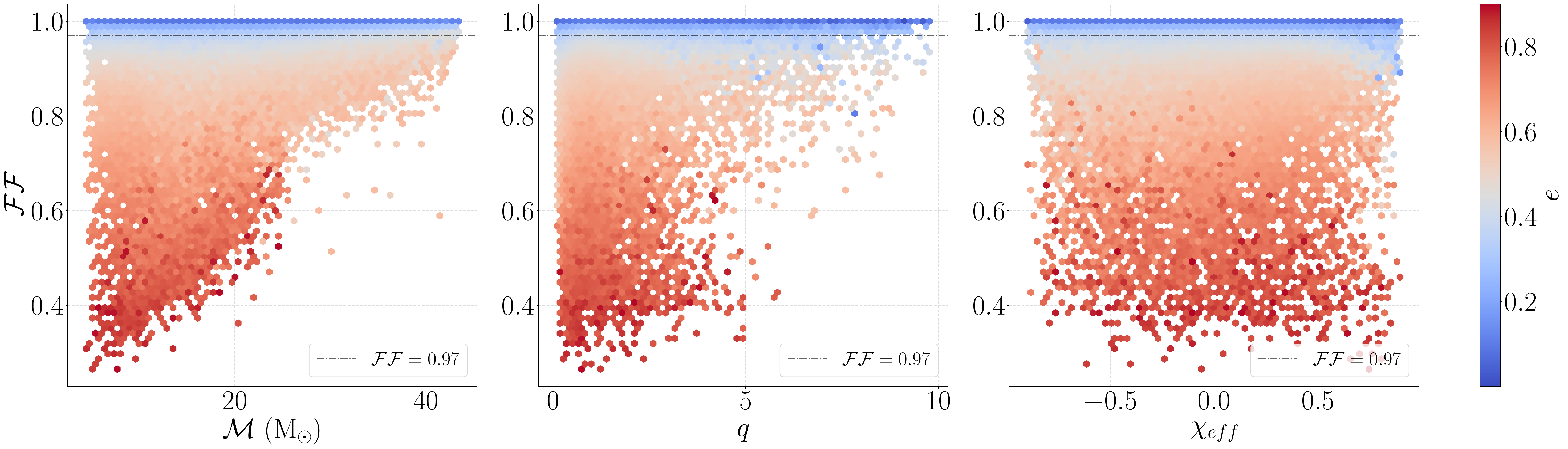}
    \caption{Recovered continuous \ff for \texttt{SEOBNRv4E} injections against \texttt{SEOBNRv4\_ROM} as a function of injected chirp mass, mass ratio, and effective spin (from left to right).}
    \label{fig:ff_mc_q_chie_plot}
\end{figure*}

\begin{figure*}[htb]
    \centering
    \includegraphics[width=\textwidth]{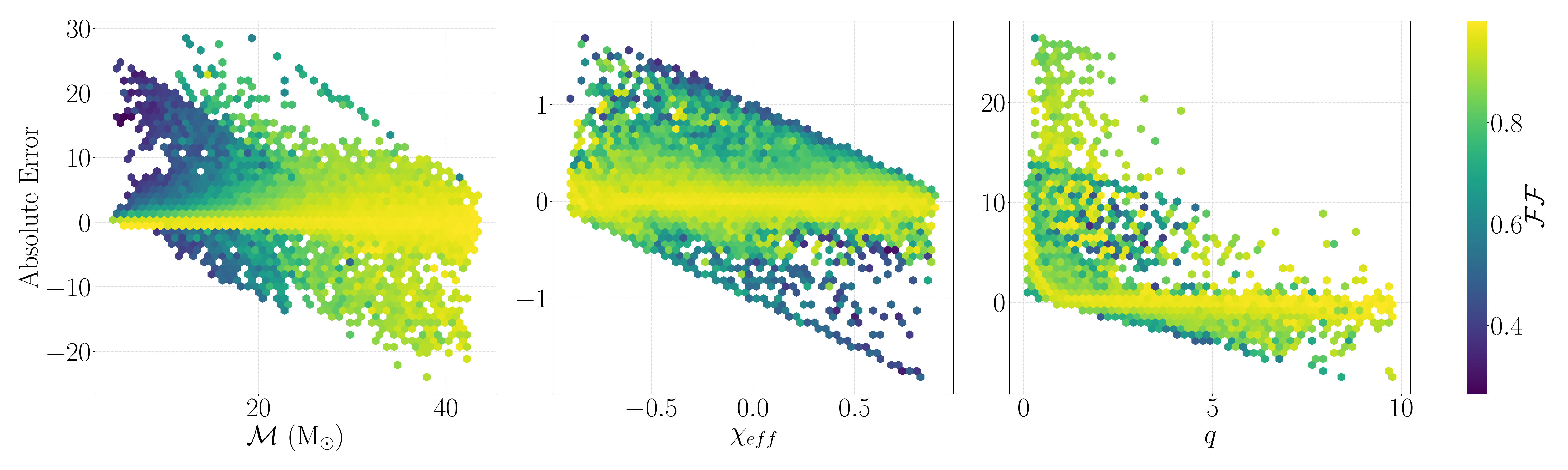}
    \caption{Plots show absolute errors as a function of injected chirp mass, effective spin, and mass ratio. The colors show $\mathcal{FF}$. Large $\mathcal{FF}$ implies smaller errors in the recovered parameters.}
    \label{fig:ff_err_mc_q_chie_plot}
\end{figure*}

\begin{figure*}[htb]
    \centering
    \includegraphics[width=\textwidth]{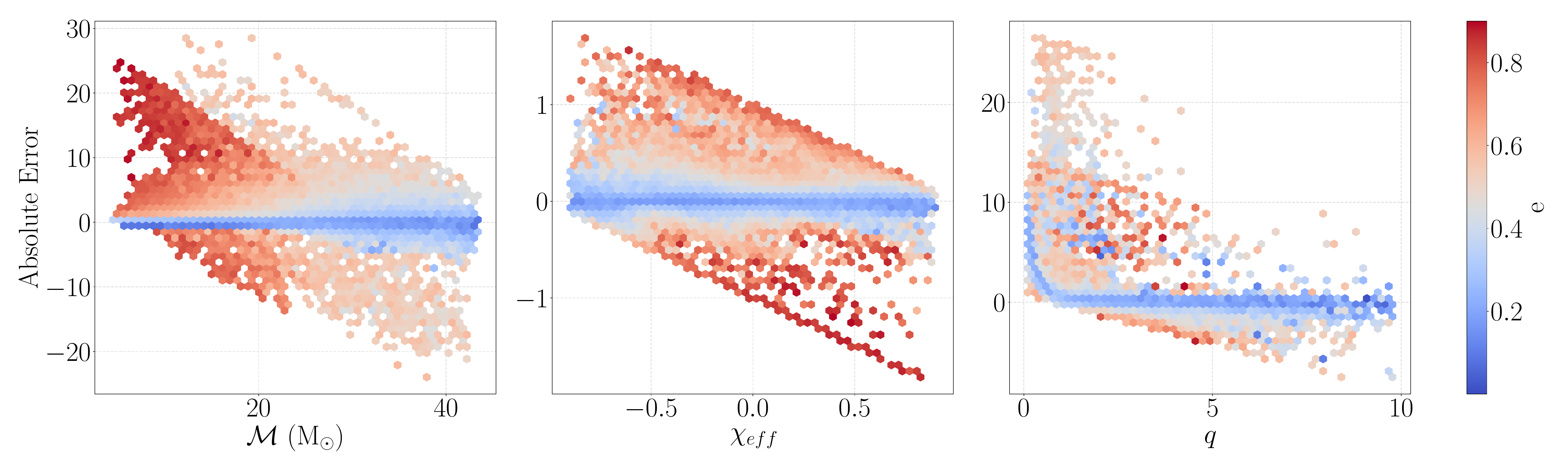}
    \caption{The figure is the same as Fig.~\ref{fig:ff_err_mc_q_chie_plot} except the colors show eccentricity.}
    \label{fig:ecc_err_mc_q_chie_plot}
\end{figure*}

To quantify and compare the sensitivities of \pycbc and \cwb searches, we use simulated quasicircular and eccentric \bbh merger signals. Before finalizing injections, we test the effectualness of non-precessing and quasicircular \texttt{SEBNRv4\_ROM} waveforms in recovering eccentric
\seob injections. To check the effectualness, we compute the fitting factor~\citep{Apostolatos_1995,fittingfactor} between the eccentric injections (using the \seob waveform model) and non-eccentric template bank (using the \texttt{SEOBNRv4\_ROM}~\cite{Cotesta:2020qhw,PhysRevD.89.061502,Bohe:2016gbl} model). The fitting factor,
\ff, is the best match between a normalized signal $s$ and a bank of normalized templates $h_i$, defined as

\begin{equation}
    \mathcal{FF} = \max_{m_1, m_2, s_{1z}, s_{2z}} \mathbb{M} \left(s, h_i \right)\,,
\end{equation}

where

\begin{equation}
    \mathbb{M} = \max_{\phi_c, t_c} \mathcal{O} \left( s, h_i \right) = \max_{t_c} \left| \mathcal{O} \left( s, (1 + i) \times h_i ~e^{-2\pi i f t_c}\right) \right|\,,
\end{equation}

represents the match maximized over the phase and time of coalescence. This match is determined by the overlap function ($\mathcal{O}$) for any two arbitrary time series $a$ and $b$ and is defined as

\begin{equation}
        \mathcal{O} (a, b) = \langle \hat{a}, \hat{b} \rangle = \frac{\langle a, b \rangle}{\langle a, a \rangle \langle b, b \rangle}\,.
\end{equation}

Here, $\langle\,,\rangle$ denotes the noise-weighted scalar product with a power spectral density ($S_n(f)$) and is given as

\begin{equation}
\label{eqn:FF}
        \langle a, b \rangle = 4 \mathcal{R}\int_{f_\text{min}}^{f_\text{max}}  \frac{\tilde{a}^*(f) \tilde{b}(f)}{S_n(f)} ~df \,.
\end{equation}

For this study, we generate 1000 eccentric \bbh signals with parameters distributed uniformly across mass ranges: $m_1$ ranging between $2-55 \, \msun$ and $m_2$ between $2-50 \, \msun$ in the detector frame. The dimensionless spins of the black holes varied up to 0.9. Additionally, we allowed relativistic anomaly, $\zeta$, to vary uniformly between $0-2\pi$. Injections are divided into two sets based on the eccentricity range: in the first low eccentricities set, $e_{10{\rm Hz}} = e$\footnote{The eccentricity $e$ is defined at 10 Hz unless specified otherwise.} is sampled uniformly between 0 and 0.3. The second set covered the eccentricity range from 0.3 to 0.9. To compute the \ff, we employ the template bank equivalent to the one used for a \pycbc \textit{broad} search as described in the third Gravitational-wave Transient Catalog (GWTC-3)~\cite{gwtc3_paper}. This template bank is constructed using the hybrid geometric-placement method~\cite{soumen,soumen2} incorporating templates with a total mass ranging from 2 to 500 $M_{\odot}$ and dimensionless spins up to 0.9. Additionally, templates exceeding 0.15 s were excluded from the bank. We do not restrict the bank using chirp mass ($\mathcal{M}$) or Newtonian time ($\tau_0$) window about each of the injection parameters. This is necessary as the waveform duration can change enough due to eccentricity even for a fixed chirp mass, as evident from Fig~\ref{fig:waveforms}.

The most eccentric injections give \ff values below the lower limit of 0.97 for a non-precessing and non-eccentric template bank, as seen from Fig.~\ref{fig:cumulative_plot_inj1inj2_comb}. These injections with low \ff values predominantly feature high chirp mass and low mass ratio ($q = m_{1}/m_{2} \ge 1$), and exhibit large eccentricities, as shown in Fig.~\ref{fig:ff_bank}. However, some injections with low eccentricities but high chirp mass and high mass ratio values also show poor \ff values. This motivates us to restrict the mass ratio to 10 for low eccentricities ($e < 0.3$) and 5 for high ($ 0.3 \le e \le 0.9$).

We perform a deeper \ff study to understand the chosen parameter space better. This study continuously maximizes the match between a given \seob injection and \texttt{SEOBNRv4\_ROM} waveforms across component masses and spins. To accomplish this, we employ two global optimization techniques -- particle swarm optimization (\pso) via the \texttt{PySwarms} package~\cite{Miranda2018} with swarm size of 20 and 200 iterations, and differential evolution (\diffevol) implemented in the \texttt{SciPy} library~\cite{2020SciPy-NMeth} with population size of 20 with 40 iterations with \textit{randtobest1bin} strategy which we found optimal. We then select the superior fitting factor obtained from these two optimization methods as our final \ff value. This approach allows us to thoroughly explore the parameter space and obtain the best match between the injected signals and the waveform model.

For quasicircular and eccentric \bbh injection sets, the component detector frame (redshifted) masses are sampled uniformly from $(3-50)\, \msun$ while non-precessing spin magnitudes are restricted to $0.9$. For eccentric injections, we used two injection sets with eccentricities drawn uniformly from $(0 - 0.3)$ and $(0.3 - 0.6)$, respectively. For each of the O3a and O3b data chunks as described in Sec.~\ref{subsec:data}, to have higher statistics and hence enable a deeper understanding of how the matched-filter search performs as a function of various source parameters \pycbc performs around 20000 injections per injection sets, whereas \cwb performs around 10000 injections per injection set. In addition to the two eccentricity ranges, both the searches also use similar sets of injection sets but with the inclusion of higher harmonics available within the \seobhm waveform model.

For eccentricities in the range $(0.6-0.9)$, we have restricted \bbh component masses to $30\, \msun$ since the merger-ringdown part of the \seob waveform model is quasicircular by construction, while high-mass eccentric BBH with high initial eccentricity can no longer be expected to circularize before the merger. Further, we distribute these injections uniformly within a chirp distance of 5 to 300 Mpc.

The findings of our \ff study are shown in Fig.~\ref{fig:ff_mc_q_chie_plot}. The x-axis of each subplot shows injected values of chirp mass, mass ratio, and effective spin, respectively, while the y-axes show \ff. The injection eccentricity goes from smaller to larger as the color changes from blue to red. When eccentricities are smaller than 0.2, \ff values are greater than 0.97, corresponding to the typical min-match criterion used in template bank constructions. However, with increasing eccentricity, \ff values go significantly lower than 0.97, indicating a severe loss in the search sensitivity.
In the eccentricity range $(0.6-0.9)$ at the reference frequency of 10 Hz, there are fewer samples than in lower eccentricity ranges because the \seob waveform model does not generate waveforms when the initial separation is less than $6/M$ (where $M$ is the total mass of the system). This is because, at these high eccentricities and small separations, the eccentricity at the attachment time of the merger-ringdown part is non-negligible, which breaks the underlying assumption of circularization of the binary before the merger and can cause unphysical features in the waveforms~\cite{Ramos-Buades:2023yhy}.

Figure~\ref{fig:ff_err_mc_q_chie_plot} and Fig.~\ref{fig:ecc_err_mc_q_chie_plot} show absolute errors in the recovered parameters as a function of injected parameters with the colors indicating \ff and eccentricities, respectively, similar to the Fig.~\ref{fig:ff_mc_q_chie_plot}. The plots show that errors in all the parameters grow larger with eccentricity while \ff values plummet. Absolute errors in the chirp mass can be larger than $100\, \%$, and the errors in mass ratio go well beyond the injected values. This indicates that using a non-eccentric template bank only covering the injected mass range may make our search based on templates even less effective.

\subsection{\lvk open data}\label{subsec:data}
In this section, we describe the LVK open data used to simulate searches for eccentric \bbh using the injection set described in Sec.~\ref{subsec:injections}.

The strain data recorded by the Advanced LIGO and Virgo detectors during their third observing run are made publicly available by LVK collaboration through the Gravitational Wave Open Science Center (GWOSC)~\cite{KAGRA:2023pio,LIGOScientific:2019lzm}. We choose two stretches of the strain data from the first and second parts of the O3 run, each roughly one week long. The first stretch of data from the first part of O3 corresponds to the GPS interval from 1251349051 till 1252015046 (UTC Interval 2019-09-01 04:57:13 - 2019-09-08 21:57:08), and the other is from the second part of O3 corresponding to GPS Interval from 1265132995 till 1265747453 (UTC Interval 2020-02-07 17:49:37 - 2020-02-14 20:30:35). The first of these data segments does not contain any known significant \bbh merger but the later stretch of data contains \textsc{GW200208\_130117}~\cite{gwtc3_paper} which we ignore in our current analysis.
In our study, we also use auxiliary information provided by the detector characterization team within LVK collaboration to analyze the times around the poor-quality data and hardware injections. In our search sensitivity analysis, we only use data jointly passing flags {\sc CBC\_CAT1},   {\sc BURST\_CAT1},  {\sc NO\_CBC\_HW\_INJ} and  {\sc NO\_BURST\_HW\_INJ}.

Below, we give details of our search pipeline configurations used to search for the injections through the O3 data.

\subsection{Matched filter search: \pycbc}
\label{subsec:pycbc_config}

 As the binary parameters are unknown, the matched-filter pipeline \pycbc uses a template bank to search for \cbc in the strain data over the parameter space of interest. In this work, we have used the offline \pycbc search configuration employed during the O3 run with the LIGO-Virgo detectors~\cite{gwtc3_paper}, albeit with the reduced parameter space coverage due to the smaller bank. We chose the restricted template bank to correctly quantify eccentricity effects missing from the non-precessing search.

As we have focused on \bbh mergers, our stochastic template bank is designed to recover \bbh with detector frame (redshifted) component masses in range $(2.5, 55)\, \msun$ with non-precessing spins amplitudes in range ($-0.998, 0.998$). Also, as we are investigating the detection capabilities of standard matched-filter searches, we are using a template bank applicable for quasicircular mergers, i.e., zero eccentricity, with the bank built from the non-precessing non-eccentric \texttt{SEOBNRv4\_ROM}~\cite{Cotesta:2020qhw,PhysRevD.89.061502,Bohe:2016gbl} waveform model. It has already been demonstrated that for small values of eccentricity, the quasicircular searches are effective over the broader parameter range, like low mass ($M < 30 \msun$) \cbc~\cite{Divyajyoti:2023rht,Huerta:2013qb,Cokelaer:2009hj,Brown:2009ng,Tessmer:2007jg} and heavier \bbh mergers~\cite{Divyajyoti:2023rht,Ramos-Buades:2020eju,Ramos-Buades:2021adz,Chen:2020lzc}.
Also, previous studies have compared the sensitivity of matched filter and morphology-independent searches using a limited set of \NR waveforms~\cite{Ramos-Buades:2020eju}. But here, we are presenting a systematic injection study covering the broad \bbh parameter space including spins, using the state-of-the-art semi-analytic eccentric IMR waveform model \seob that includes spin-eccentricity cross terms.

\subsection{Un-modelled search : \cwb}
\label{subsec:cwb_config}
\cwb is a morphology-independent analysis pipeline that detects and reconstructs GW signals without assuming any waveform model~\cite{klimenko2016method, klimenko_sergey_2021_5798976, drago2021coherent}. \texttt{cWB} decomposes each interferometer data into a time-frequency representation using Wilson-Daubechies-Meyer wavelets~\cite{necula2012transient}.
Each wavelet amplitude is normalized by the corresponding detector amplitude spectral density, \texttt{cWB}, then selects those wavelets having energy above a fixed threshold. Finally, clusters from different detectors are combined coherently into a likelihood function, which is maximized with respect to the sky location. \texttt{cWB} is a versatile algorithm that is used to search for a wide variety of transient GW sources like direct captures and hyperbolic encounters of BBHs~\cite{Ebersold:2022zvz,Bini:2023gaj}, non-linear memory~\cite{Ebersold:2020zah} and generic transients~\cite{KAGRA:2021tnv,KAGRA:2021bhs}. The instance of \texttt{cWB} employed here is the one that is tuned to detect stellar mass BBHs. In this instance of \cwb, the excess energy threshold above the noise floor of the detectors on the time-frequency cluster is chosen to be a diagonal pattern, which mimics broadly the chirp signal. The post-production cuts deployed to mitigate the background trigger are the same as those used for searching BBHs in GWTC-3~\cite{gwtc3_paper}. It should be noted that we have not used the post-production cuts of \cwb which was deployed by the LVK collaboration for the search of eBBHs during the third observing run, as the post-production cuts used there are tailored towards searching eBBHs at higher masses~\cite{lvk_ebbh_2023}. In this study, we focus on the low chirp mass systems; hence, the \cwb used in GWTC-3 is deployed here.

\section{\label{sec:results}Results}

\subsection{Search sensitivity}

To quantify search sensitivity, we estimate the sensitive volume of each of the search pipelines. The sensitive volume of the search is proportional to the efficacy of the search with simulated signals. The efficacy of the search is the ratio of found injections to the total injections and hence is a function of detection threshold and injection distribution. We use injections distributed uniformly in component masses and chirp distance, following the procedure detailed in Sec. 4.1 of Ref.~\cite{Usman:2015kfa} to compute the sensitive volume.

We search through strain data to detect simulated injections using \pycbc and \cwb pipelines. We label the injection as found if it is detected by the search pipeline with a conservative \far $< 1$ per $2$ yr. In the following sections, we describe and compare the measured sensitivities for each of the searches for eccentric (as well as quasicircular) \seob and \seobhm injections up to eccentricity $< 0.6$. In the following sections, unless specified otherwise, we discuss the combined results of \seob and \seobhm injections.

\subsection{PyCBC results}

\begin{figure*}[htb]
    \begin{minipage}{.5\textwidth}
        \includegraphics[width=\textwidth]{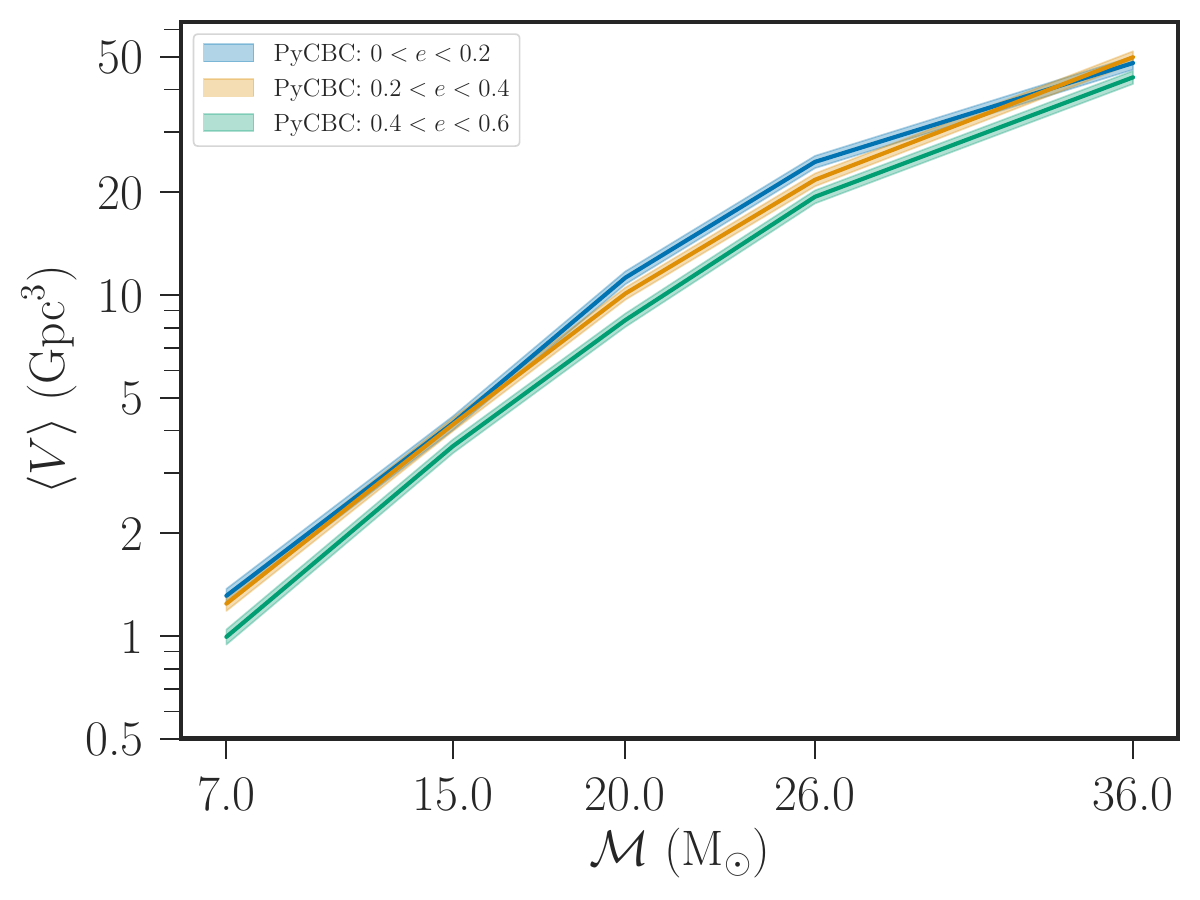}
        \caption{Average sensitive search volume as a function of chirp mass for different eccentricity bins. Sensitivity decreases with increasing eccentricity for all the chirp mass bins.}
        \label{fig:py_mc_e_v}
    \end{minipage}%
    \begin{minipage}{.5\textwidth}
        \includegraphics[width=\textwidth]{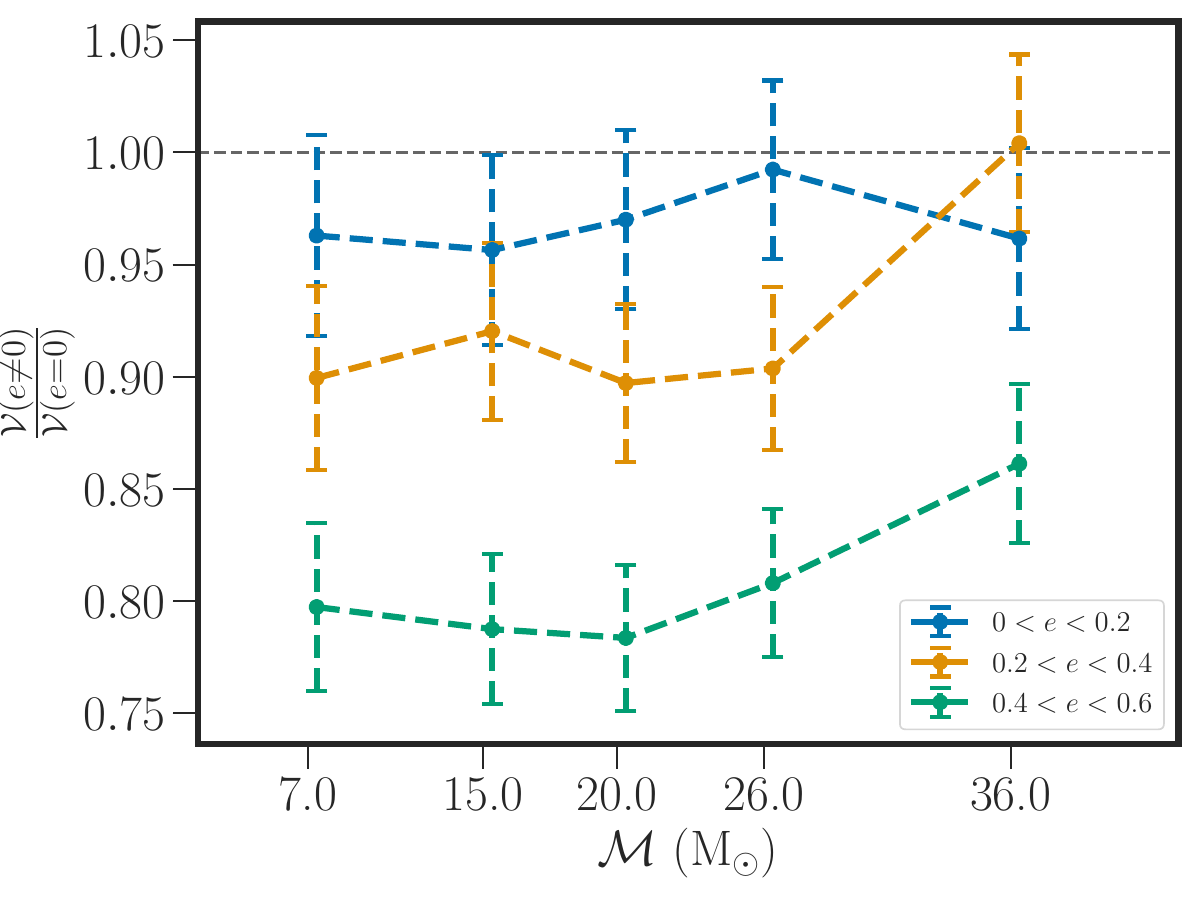}
        \caption{Ratio of sensitivity volumes for eccentric to circular \bbh injections as a function of chirp mass. }
        \label{fig:py_mc_e_v_ratio}
    \end{minipage}
\end{figure*}

\begin{figure*}[htb]
    \begin{minipage}{.5\textwidth}      \includegraphics[width=\textwidth]{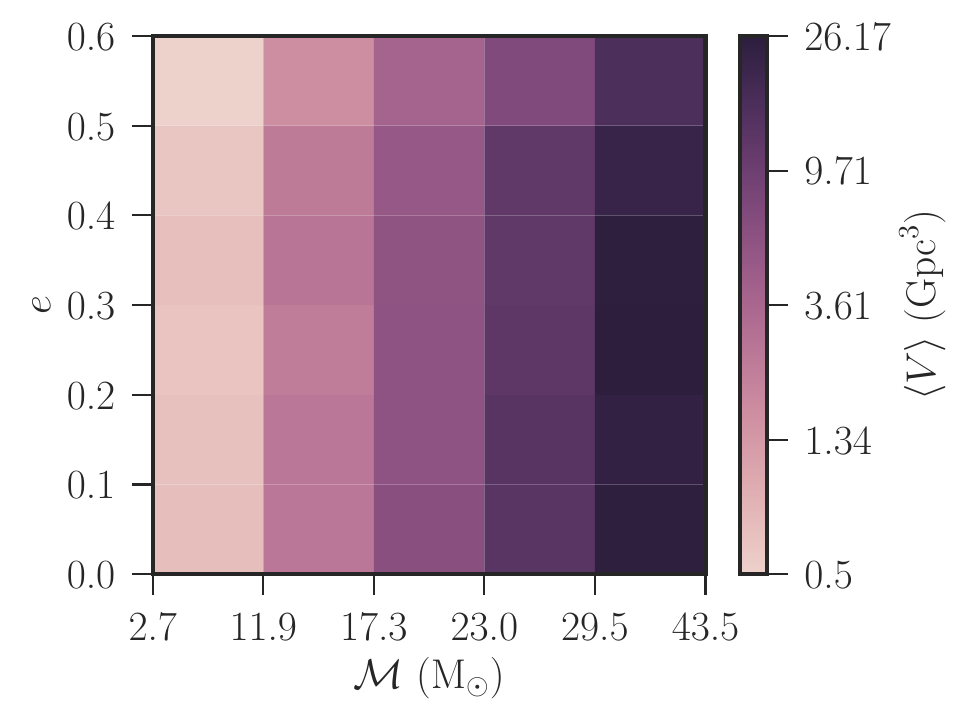}
        \caption{Average sensitive volume of the search as a function of chirp mass and eccentricity. Sensitivity increases with chirp mass and decreasing eccentricity. This is another representation of Fig.~\ref{fig:py_mc_e_v}. }
        \label{fig:py_mc_e_v_image}

     \end{minipage}%
    \begin{minipage}{.5\textwidth}
        \includegraphics[width=\textwidth]{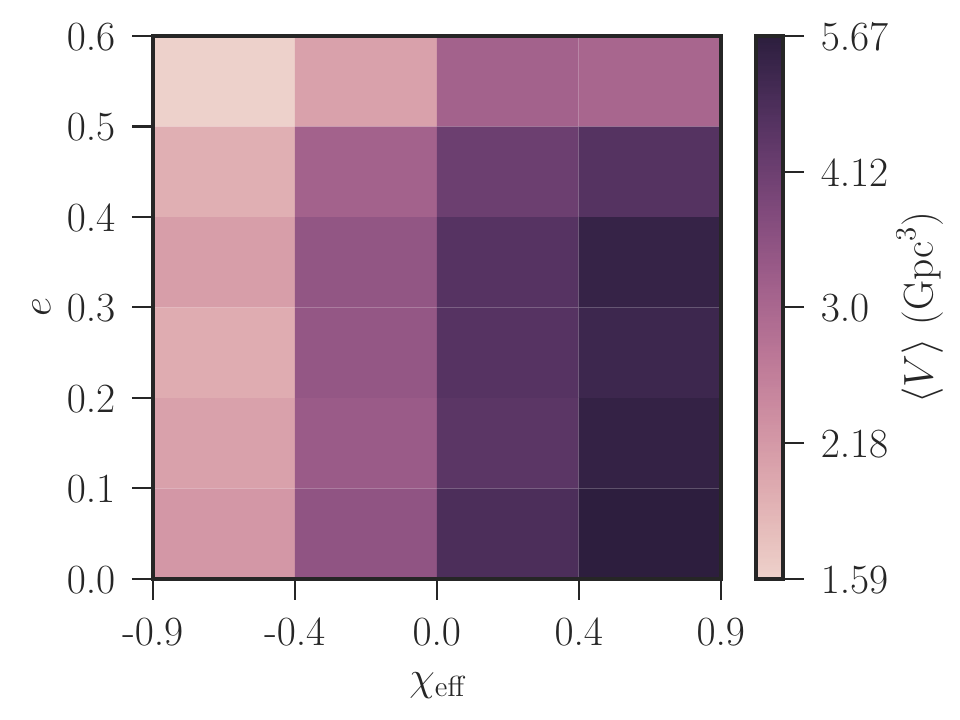}
        \caption{Similarly, average sensitivity of the search as a function of eccentricity and effective spin parameter. }
        \label{fig:py_e_chie_v}
    \end{minipage}
\end{figure*}

\begin{figure*}[htb]
    \begin{minipage}{.5\textwidth}
        \includegraphics[width=\textwidth]{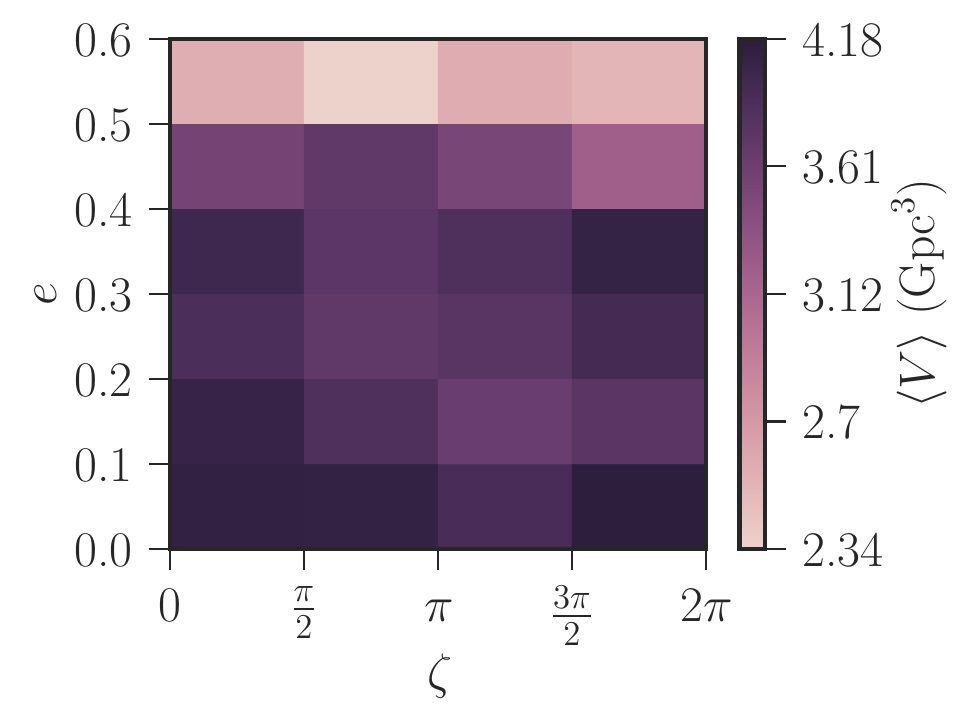}
        \caption{Plot shows sensitivity as a function of relativistic anomaly and eccentricity}
        \label{fig:py_e_l_v}
    \end{minipage}%
    \begin{minipage}{.5\textwidth}
        \includegraphics[width=\textwidth]{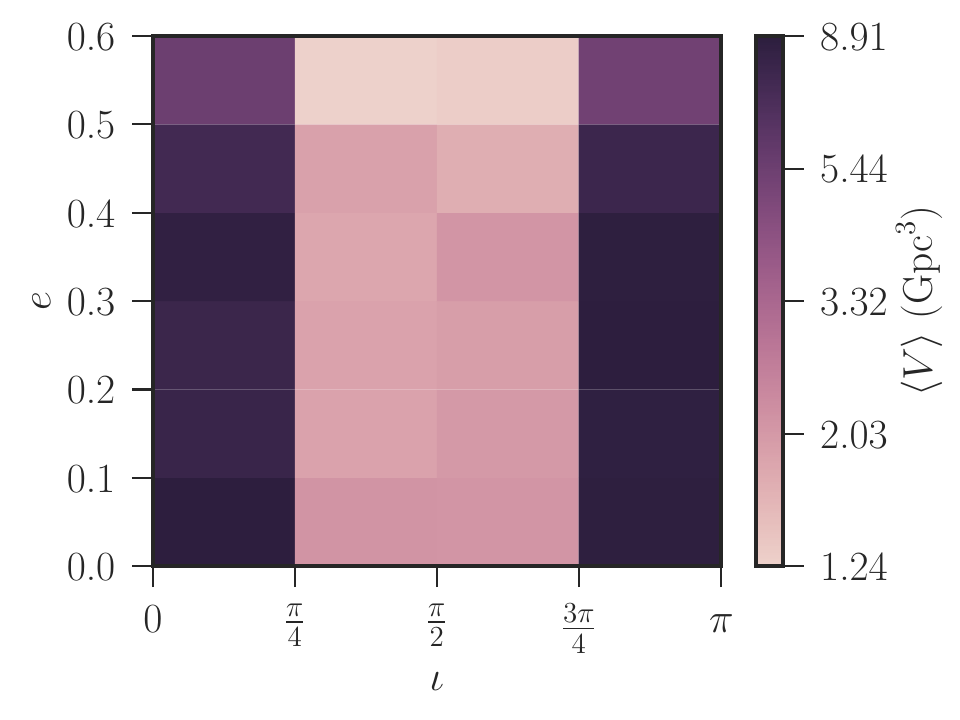}
        \caption{Sensitive volume as a function of inclination angle and eccentricity.}
        \label{fig:py_e_i_v}
    \end{minipage}
\end{figure*}

\begin{figure*}[htb]
    \begin{minipage}{.5\textwidth}
        \includegraphics[width=\textwidth]{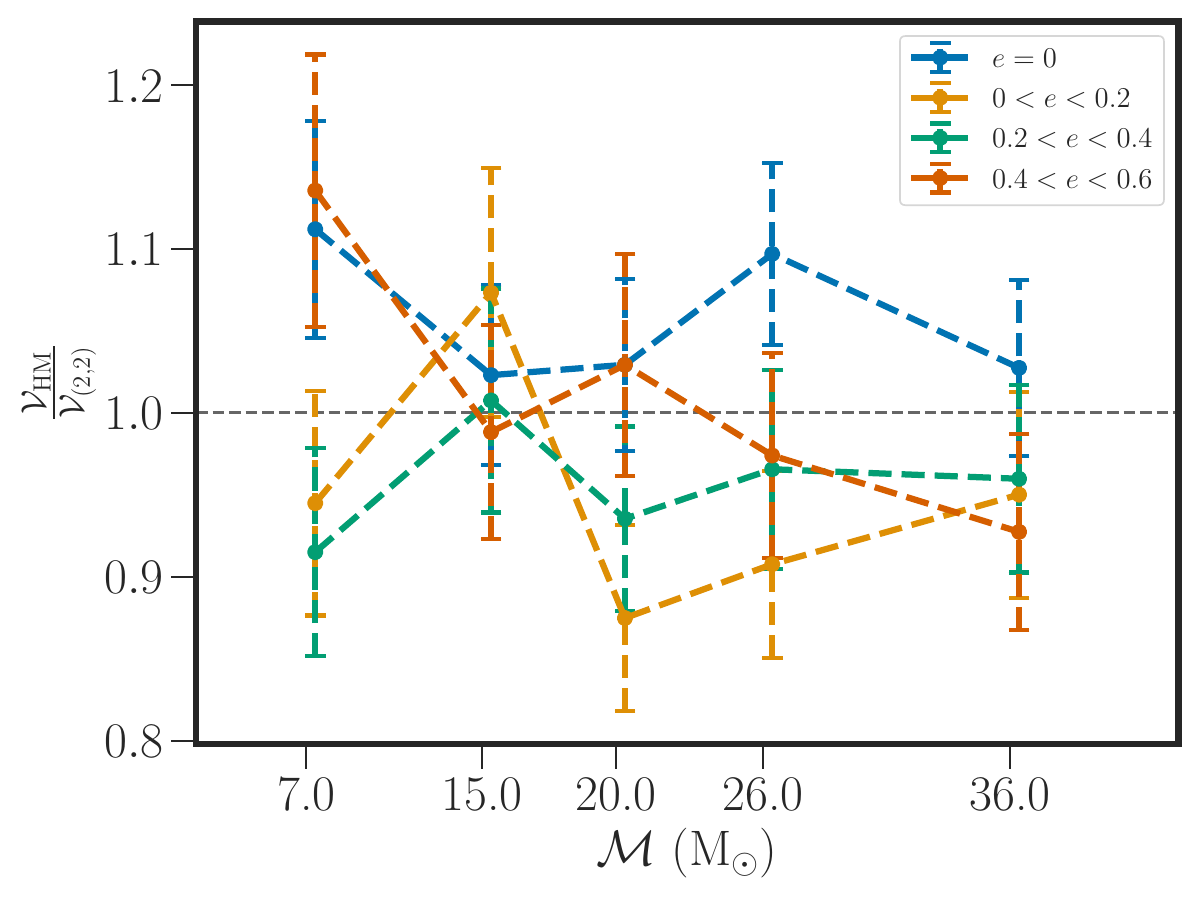}
        \caption{The ratio of the search sensitivities for eccentric BBH injections with higher modes against only dominant mode as a function of chirp mass and eccentricity.}
        \label{fig:py_mc_vratio_hm22_e}
    \end{minipage}%
    \begin{minipage}{.5\textwidth}
        \includegraphics[width=\textwidth]{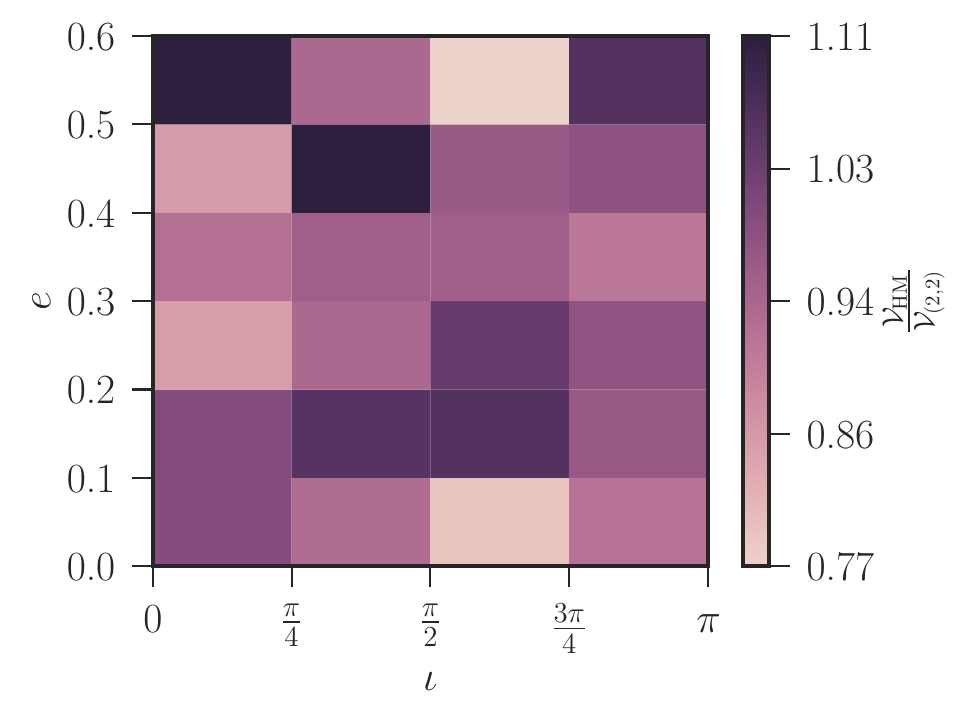}
        \caption{The ratio of the search sensitivities for eccentric BBH injections with higher modes against dominant mode. }
        \label{fig:py_e_i_vratio_hm22}
    \end{minipage}
\end{figure*}

\begin{figure*}[htb]
    \centering
    \begin{minipage}{.5\textwidth}
        \includegraphics[width=\textwidth]{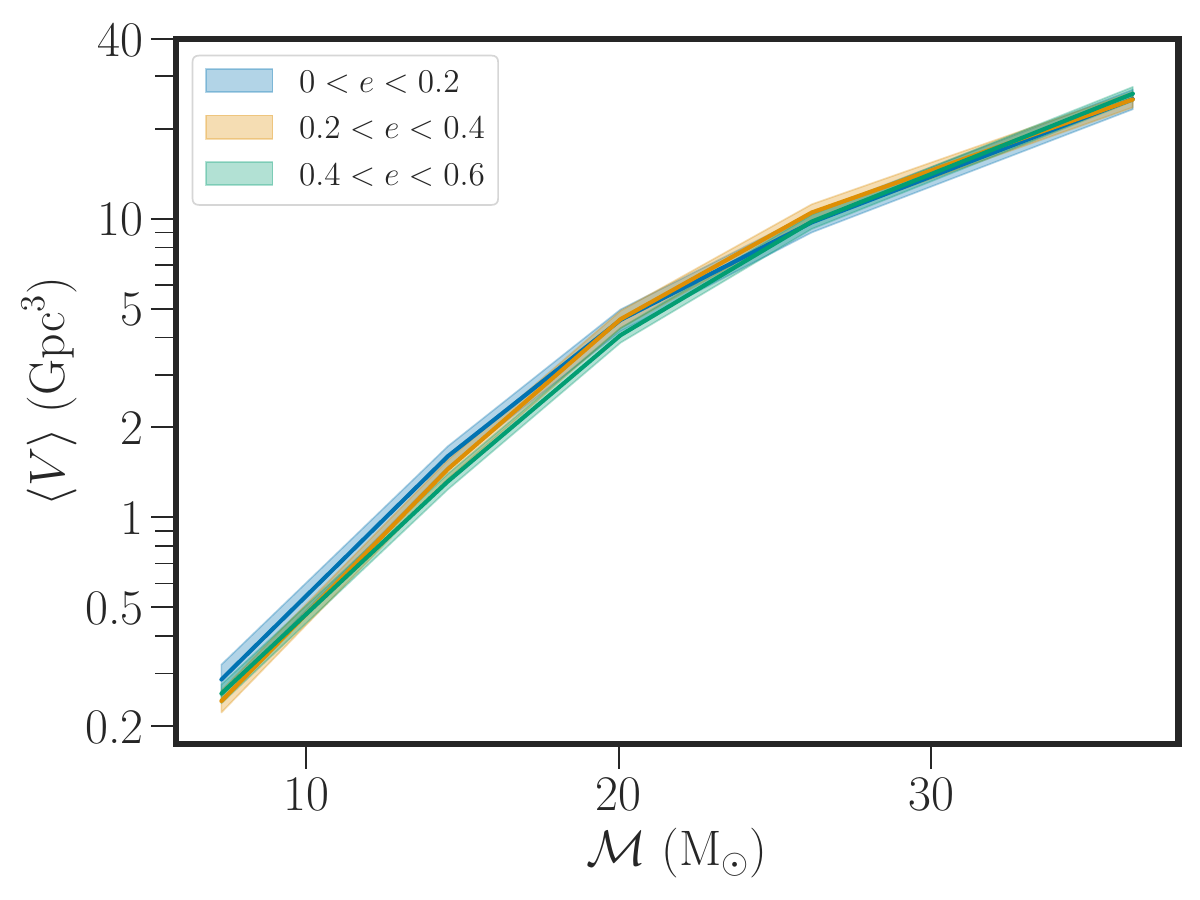}
        \caption{Average sensitive volume of the unmodelled burst search as a function of chirp mass and eccentricity.}
        \label{fig:cwb_mc_e_v}
    \end{minipage}%
    \begin{minipage}{.5\textwidth}
        \includegraphics[width=\textwidth]{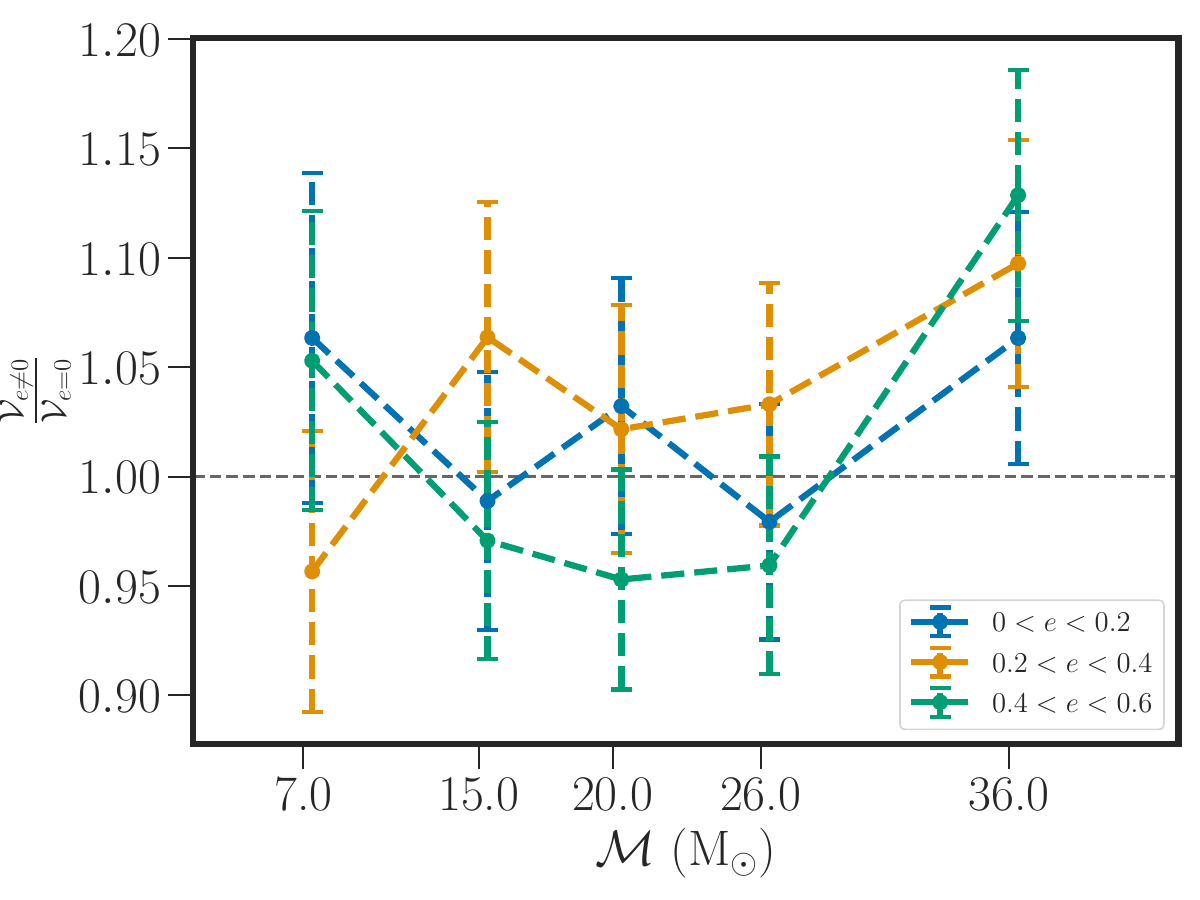}
        \caption{Ratio of sensitivity volume for eccentric to circular BBHs injections as a function of chirp mass for \texttt{cWB} search.}
        \label{fig:cwb_mc_e_v_ratio}
    \end{minipage}
\end{figure*}

\begin{figure*}[htb]
    \begin{minipage}{.5\textwidth}
        \includegraphics[width=\textwidth]{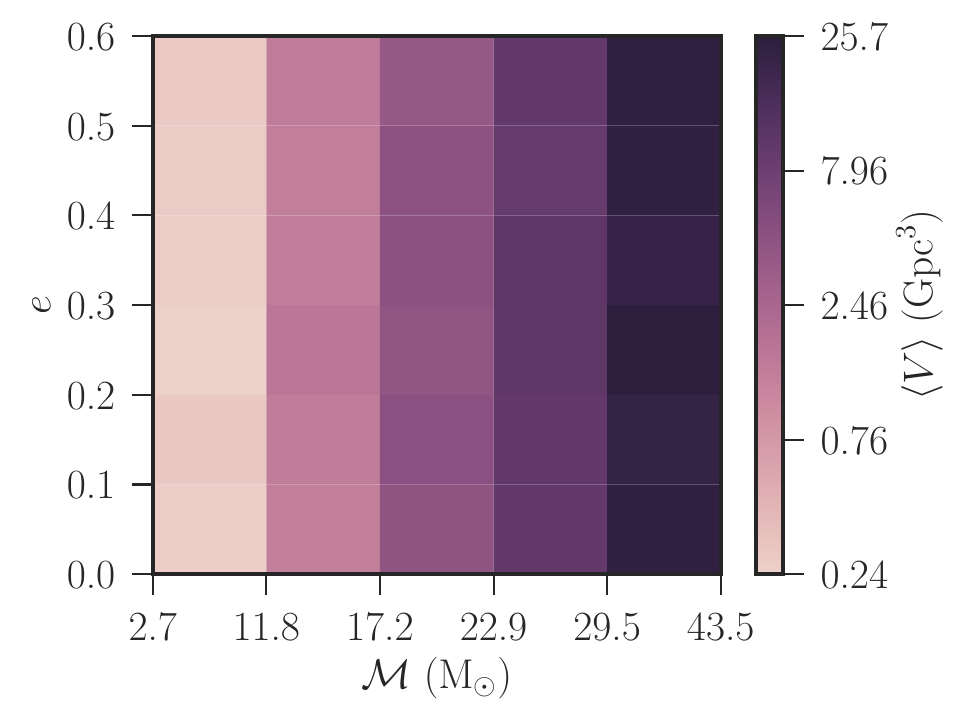}
        \caption{Average sensitive search volume as a function of chirp mass and eccentricity. This is another representation of Fig.~\ref{fig:cwb_mc_e_v}.}
        \label{fig:cwb_mc_e_v_image}
    \end{minipage}%
    \begin{minipage}{.5\textwidth}
        \includegraphics[width=\textwidth]{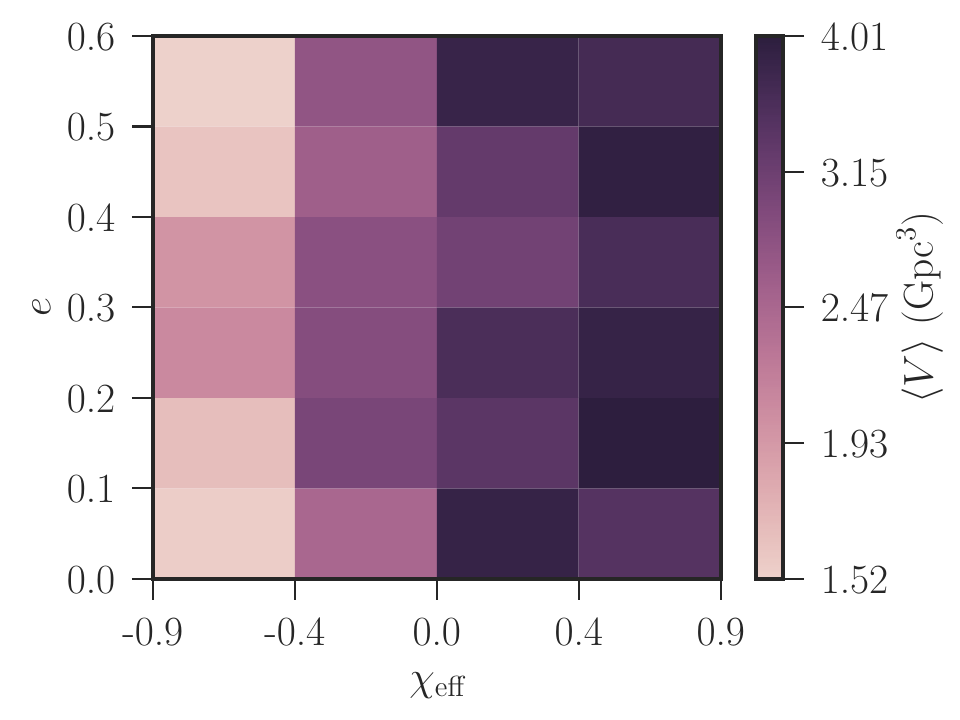}
        \caption{The average sensitivity of \texttt{cWB} as a function of eccentricity and effective spin.}
        \label{fig:cwb_mc_e_chie_v}
    \end{minipage}
\end{figure*}

\begin{figure*}[htb]
    \begin{minipage}{.5\textwidth}
        \includegraphics[width=\textwidth]{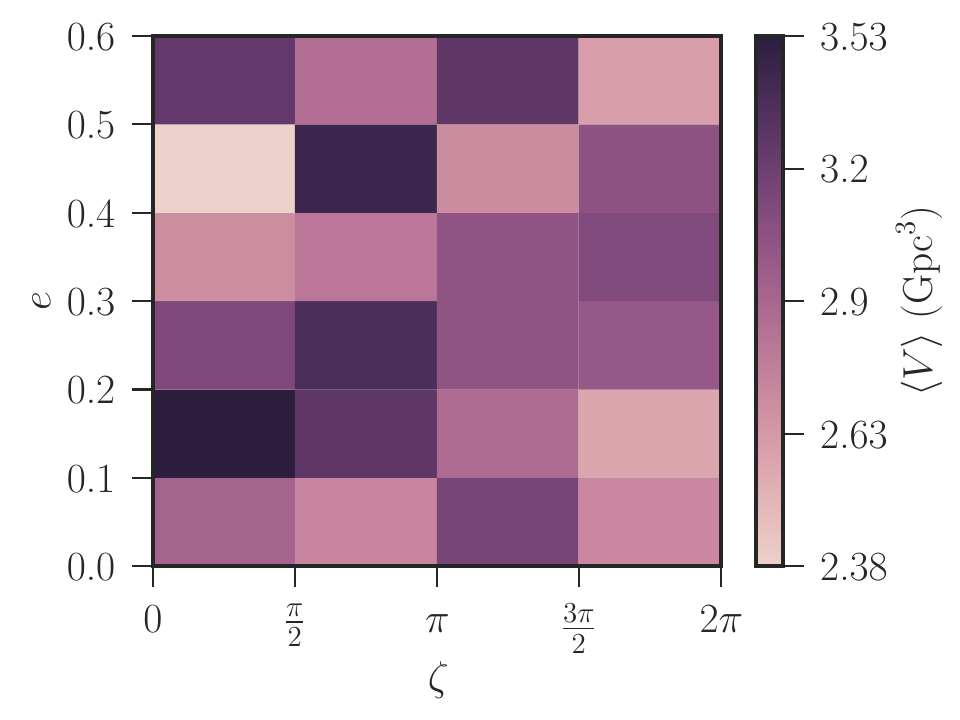}
        \caption{Average sensitive volume as a function of relativistic anomaly and eccentricity.}
        \label{fig:cwb_e_l_v}
    \end{minipage}%
    \begin{minipage}{.5\textwidth}
        \includegraphics[width=\textwidth]{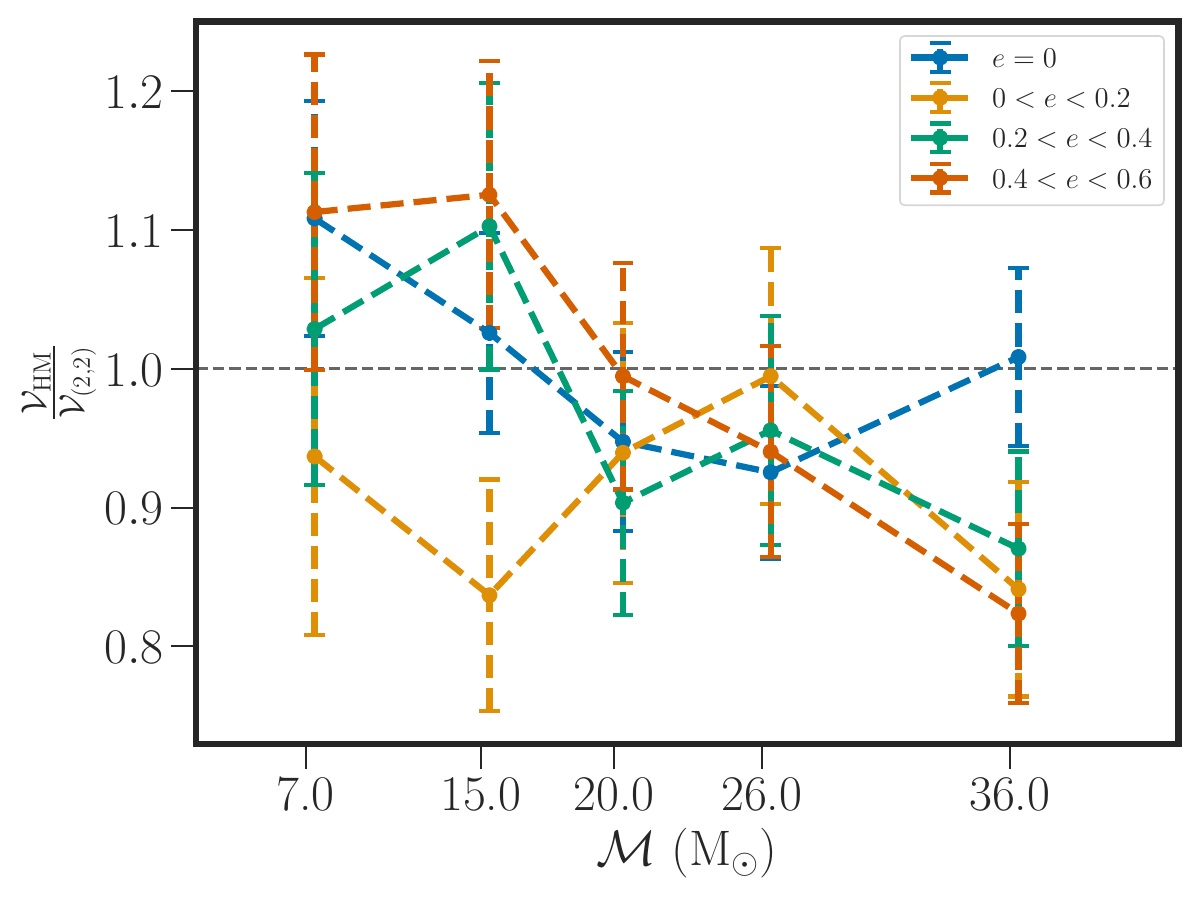}
        \caption{The ratio of sensitive volumes of \cwb for BBH injections with higher modes against only the dominant mode as a function of eccentricity and chirp mass bins.}
        \label{fig:cwb_mc_vratio_hm22_e}
    \end{minipage}
\end{figure*}

For the \pycbc search configuration described in Sec.~\ref{subsec:pycbc_config}, we estimated sensitive volume using various subsets of injections. Figure~\ref{fig:py_mc_e_v} shows sensitive volume (with $1 \sigma$ errors) as a function of injected chirp mass. For the chirp mass bin of $10 \, \msun$, sensitivity is more than 1 Gpc$^3$ and goes to 6 Gpc$^3$ around the chirp mass of $20\, \msun$. The search loses sensitivity across the mass range with increasing eccentricity. The ratio of sensitive volumes of the eccentric bins against quasicircular \bbh injections is plotted in Fig.~\ref{fig:py_mc_e_v_ratio}. For eccentricities in the range $(0-0.2)$, the sensitivity loss is at best, a few percent. But injections with eccentricities in the range $(0.2 - 0.4)$ and $(0.2 - 0.4)$ show $5-10\,\%$ and $15-20\, \%$ loss in sensitivity across the mass range with maximum sensitivity is lost for chirp mass around $20 \, \msun$ as the LIGO detectors are most sensitive in that mass region. This is true even though eccentric \bbh mergers are expected to be more luminous~\cite{Chen:2020lzc} than quasi-circular mergers.

Figure~\ref{fig:py_mc_e_v_image} is just another representation of Fig.~\ref{fig:py_mc_e_v} with finer grid in eccentricity and without error bars. Colors from lighter to darker indicate the expected sensitivity pattern of increasing volume with chirp mass and decreasing eccentricity. A similar trend is seen in Fig.~\ref{fig:py_e_chie_v} with the effective spin parameter ($\chi_{\rm eff}$). When spins are aligned with the orbital angular momentum, the search sensitivity is highest, while it is lowest when the spins are anti-aligned.

Furthermore, we see in Fig.~\ref{fig:py_e_l_v} that relativistic anomaly does not affect the search sensitivity compared to the eccentricity. As expected, \pycbc can detect face-on or face-off binary orientations farthest compared to edge-on cases, as seen in Fig.~\ref{fig:py_e_i_v}. The presence of higher modes affects search sensitivity for the injection population we have considered but in an unpredictable pattern (Fig.~\ref{fig:py_mc_vratio_hm22_e}). For quasicircular systems, injections with higher modes show $5-10\, \%$ sensitivity gain (blue dashed line). But with injections having eccentricities in the range of ($0.2-0.4$), the search loses $5-10\, \%$ of sensitive volume due to the presence of higher harmonics (green dashed line). The other two eccentricity bins show relative gain in the sensitivity for lower chirp masses, but there is a reduction in the search-sensitive volume for larger chirp mass bins, as compared to only having the dominant mode. Similarly, the ratio of sensitive volumes with and without higher modes also does not show any pattern as a function of eccentricity and inclination, as seen in Fig.~\ref{fig:py_e_i_vratio_hm22}.

All these results signify that we do need a dedicated search to detect \bbh mergers with eccentricities ($e_{10{\rm Hz}}$) larger than 0.2.  We also note that for mass ratios smaller than 5, the effect of higher harmonics on the template-based search is limited to a few per cent.

\subsection{cWB results}

The sensitivity of \cwb increases as a function of chirp mass, going from $\sim 0.4$ Gpc$^3$ around chirp mass of $10 \, \msun$ to $\sim 35$ Gpc$^3$ for chirp mass of about $35 \, \msun$, as shown in Fig.~\ref{fig:cwb_mc_e_v}. However, the sensitive volume estimated for \cwb is largely unaffected by the presence of eccentricity. This can be seen in Fig.~\ref{fig:cwb_mc_e_v_ratio}, which shows the ratio of sensitive volume for eccentric injections compared to quasicircular. The volume ratio fluctuates around unity with error bars as large as $10\, \%$ except for the heaviest chirp mass bin where the \cwb search shows $5-10\, \%$ more sensitivity to eccentric injections than quasicircular ones.

Figure~\ref{fig:cwb_mc_e_v_image} shows the sensitivity of the burst search as a function of chirp mass and eccentricity. Similar to the curves in Fig.~\ref{fig:cwb_mc_e_v}, the image shows sensitivity increases with chirp mass but is unaffected by the eccentricity. Similar to the template bank-based search, \cwb can detect \bbh with large spins aligned along the orbital angular momentum more than twice as far as compared to anti-aligned spins (Fig.~\ref{fig:cwb_mc_e_chie_v}). The sensitivity of the burst search is also largely unaffected by the relativistic anomaly, similar to the eccentricity, evident from Fig.~\ref{fig:cwb_e_l_v}.

While comparing the respective sensitivities of the burst search to \seobhm injections and \seob injections in Fig~\ref{fig:cwb_mc_vratio_hm22_e}, we find that the search is a few per cent more sensitive for \bbh with chirp mass lighter than $15\, \msun$ when eccentricity is fairly large $(>0.2)$ (dashed green and orange lines). Interestingly, the search loses $10\, \%$ of its sensitive volume for very massive eccentric $(\mathcal{M} > 26\, \msun)$ \bbh mergers but shows similar sensitivity even with the presence of higher harmonics for heavy quasicircular mergers (dashed blue line). With the reduction in the signal duration with increasing eccentricity (see Fig.~\ref{fig:waveforms}) and more mergers being more luminous~\cite{Chen:2020lzc}, one expects \cwb to perform better for eccentric \bbh mergers with higher harmonics as compared to the quasicircular mergers. This is not evident for the low chirp mass parameter space as \cwb's sensitivity for BBHs is mostly driven by the merger part of the signal. This also provides the direction towards which the clustering procedure of time-frequency pixel for eBBH in \cwb search can be improved.

\subsection{Comparison}

\begin{figure*}[htb]
    \centering
    \begin{minipage}{.5\textwidth}
        \includegraphics[width=\textwidth]{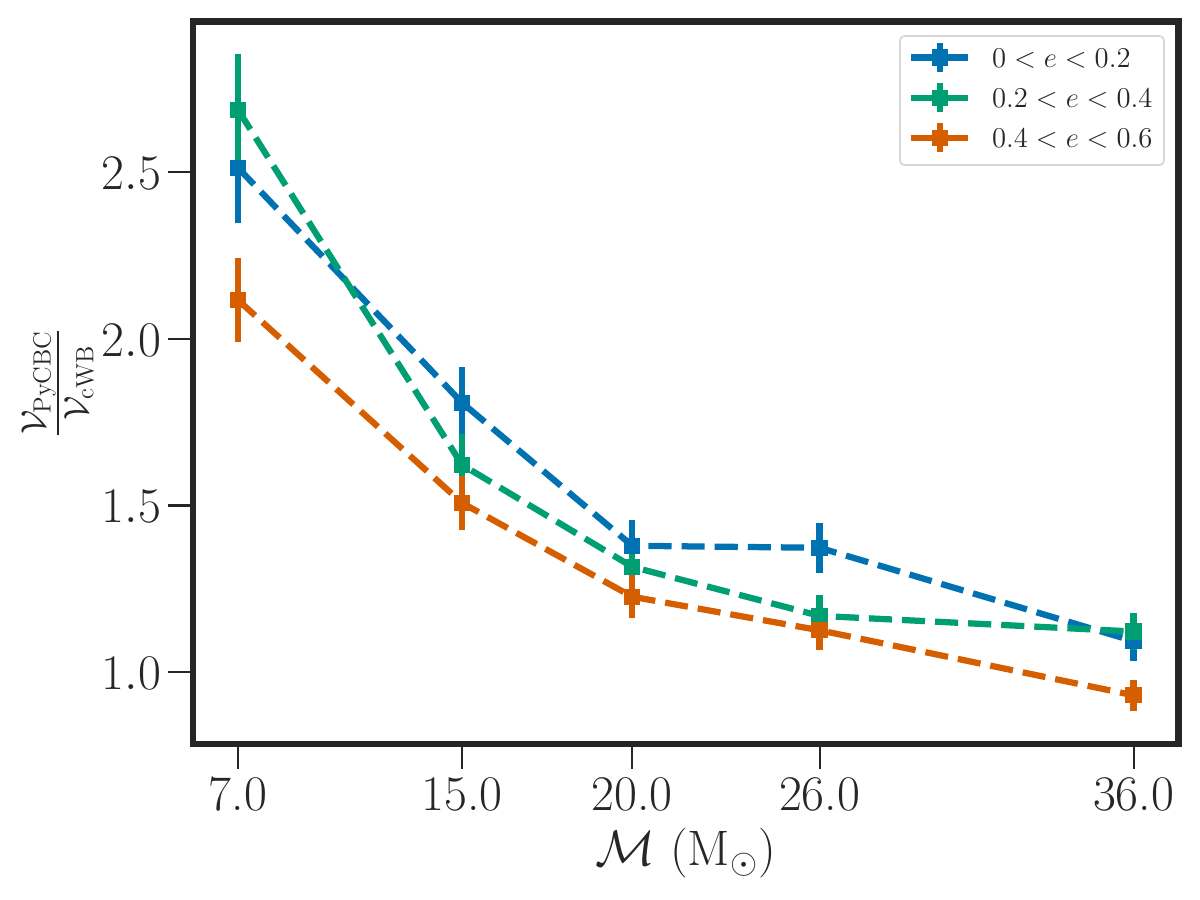}
        \caption{The plot shows the ratio of both the search sensitivities as a function of chirp mass. \pycbc is more than twice as sensitive as \cwb when chirp mass is less than 10 $\msun$, but \cwb catches up as chirp mass increases.}
        \label{fig:py_vs_cwb_vratio_mc_e}
    \end{minipage}%
    \begin{minipage}{.5\textwidth}
        \includegraphics[width=\textwidth]{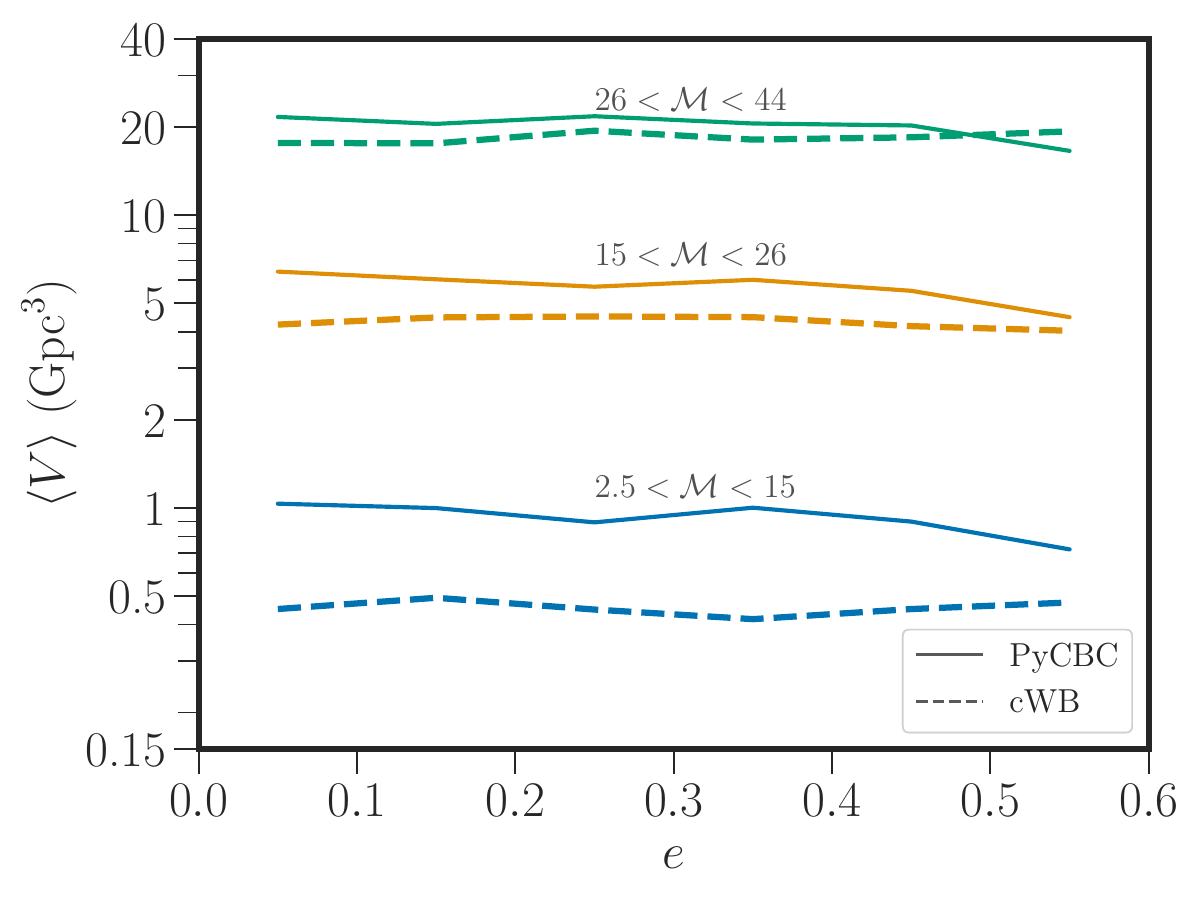}
        \caption{Average search sensitivity as a function of eccentricity for three different chirp mass bins. Solid lines represent \pycbc and dashed corresponds to \cwb. It is clear that with increasing chirp mass and }
        \label{fig:py_vs_cwb_v_e_mc}
    \end{minipage}
\end{figure*}
Finally, we compare the performance of both pipelines, the matched filter-based \pycbc and the burst search \cwb. Figure~\ref{fig:py_vs_cwb_vratio_mc_e} shows the ratio of the sensitive volume of \pycbc to that of \cwb. For \bbh in the lightest chirp mass bin, \pycbc is more than twice as sensitive as \cwb. With increasing chirp mass, the sensitivity ratio approaches unity for the heaviest of the \bbh mergers. This means that both searches have similar sensitivity for merging compact binaries with chirp mass of $30\, \msun$ or heavier. In general, with increasing eccentricity, the sensitivity ratio between the searches reduces. This is a consequence of \pycbc losing sensitivity with increasing eccentricity.

Figure ~\ref{fig:py_vs_cwb_v_e_mc} is another representation of Fig.~\ref{fig:py_vs_cwb_vratio_mc_e} that shows the sensitive volume of a search as a function of eccentricity for different chirp mass bins. Sensitive volumes of \cwb and \pycbc are shown by dashed and solid lines, respectively, with chirp mass bins denoted by colors. With increasing chirp mass and increasing eccentricity, the sensitivity of \cwb approaches that of \pycbc. For the heaviest chirp mass bin with eccentricities in the range (0.5, 0.6), \cwb outperforms \pycbc (green dashed line crossing over the green solid line).

This comparison again underlines the need for both the search pipelines, as \cwb exceeds the sensitivity of \pycbc for shorter duration and poorly modeled GW signals. Also, the loss of \pycbc sensitivity with increasing eccentricity underlines the need for a dedicated matched filter-based search for eccentric \bbh mergers. Despite what is expected for shorter \bbh waveform durations, the \cwb sensitivity did not grow with eccentricity.

\section{\label{sec:conclusion}Conclusion}

In conclusion, this study systematically compared the sensitivities of two \bbh search approaches for mergers in eccentric orbits. Utilizing the \pycbc, a matched filter-based search pipeline, and the un-modelled burst search pipeline \cwb, configured similarly to the analysis conducted by the LVK collaboration during the third observation run, we explored the impact of eccentricity on detection capabilities.

Across the mass range of $5-50 \, \msun$ in the detector frame (redshifted), it became evident that neglecting eccentricity in the searches led to missed opportunities for detecting eccentric \bbh mergers. Specifically, the \pycbc approach, using quasicircular waveforms, exhibited a sensitivity volume loss of $10-20\, \%$ for \bbh mergers with eccentricities exceeding $0.2$, while maintaining $97-100\, \%$ sensitivity for eccentricities below $0.2$ at a reference frequency of 10Hz. In contrast, \cwb showed resilience to eccentricity, showing a slight increase in sensitivity volume. In particular, \cwb shows a $10\, \%$ increase in the sensitivity towards heavier ($\mathcal{M} \ge 30 \, \msun$) \bbh mergers in eccentric orbits.  As eccentricity increases, the waveforms of BBH mergers become shorter as the black holes merge rapidly while emitting more power~\cite{Chen:2020lzc}. This suggests the potential to observe these systems at greater distances. However, the sensitivity of the \cwb search pipeline does not exhibit a significant increase in volume as discussed before. This is attributed to the specific tuning of the pipeline configuration optimized for detecting the merger part of the signal.

Notably, the \seobhm waveform model includes spin-eccentricity cross terms. This is the first systematic search sensitivity study to incorporate these effects. In addition, we found that the relativistic anomaly and inclusion of higher harmonics within the injected eccentric signals do not affect the search sensitivity systematically for the injected population used.

The measuring eccentricity is one of the key discriminators for distinguishing between dynamic and isolated formation channels of \bbh mergers along with the spin and mass ratio measurements. Our \ff study proves that neglecting eccentricity will systematically bias the measurements of chirp mass, mass ratio, and effective spin parameter. This bias not only impedes our understanding of the BBH merger population~\cite{Zevin:2021rtf,Favata:2021vhw,Divyajyoti:2023rht,Zeeshan:2024ovp,Wagner:2024ecj,PhysRevD.92.044034} and its surroundings but also constrains our ability to test general relativity~\cite{Saini:2022igm,Narayan:2023vhm} and conduct precision cosmology with BBH mergers~\cite{Yang:2022fgp}.

The loss on sensitive volume is directly proportional to the number of missed eccentric \bbh mergers, prompting the necessity for a dedicated matched-filter-based search tailored for eccentric BBH mergers. Meanwhile, \cwb can be optimized to detect highly eccentric heavy \bbh mergers at the fractional cost compared to matched filter searches. This imperative development is essential to enhance our capacity to probe the diverse astrophysical scenarios leading to BBH mergers and advance our capabilities for precision measurements in gravitational wave astronomy.

\begin{acknowledgments}
B. G. is supported by the research program of the Netherlands Organization for Scientific Research (NWO). M. H. and A. R. B. are also supported by the NWO, and this publication is part of the Veni project  VI.Veni.222.396 financed by NWO. K. S. acknowledges the Inter-University Centre of Astronomy and Astrophysics (IUCAA), India, for the fellowship support during the initial phase and the National Science Foundation Awards (PHY-2309240) for the later phase of this work. ST is supported by the SNSF Ambizione Grant Number: PZ00P2-202204.

 The authors are grateful for computational resources provided by Nikhef, NWO and the LIGO Laboratory and supported by the National Science Foundation Grants No. PHY-0757058 and No. PHY-0823459.

This research used data or software from the Gravitational Wave Open Science Center (gwosc.org), a service of the LIGO Scientific Collaboration, the Virgo Collaboration, and KAGRA. This material is based upon work supported by NSF's LIGO Laboratory, which is a major facility fully funded by the National Science Foundation, as well as the Science and Technology Facilities Council (STFC) of the United Kingdom, the Max-Planck-Society (MPS), and the State of Niedersachsen/Germany for support of the construction of Advanced LIGO and construction and operation of the GEO600 detector. Additional support for Advanced LIGO was provided by the Australian Research Council. Virgo is funded through the European Gravitational Observatory (EGO), the French Centre National de Recherche Scientifique (CNRS), the Italian Istituto Nazionale di Fisica Nucleare (INFN), and the Dutch Nikhef, with contributions by institutions from Belgium, Germany, Greece, Hungary, Ireland, Japan, Monaco, Poland, Portugal, Spain. KAGRA is supported by the Ministry of Education, Culture, Sports, Science and Technology (MEXT), Japan Society for the Promotion of Science (JSPS) in Japan; National Research Foundation (NRF) and the Ministry of Science and ICT (MSIT) in Korea; Academia Sinica (AS) and National Science and Technology Council (NSTC) in Taiwan.

\end{acknowledgments}

\nocite{*}

\bibliography{apssamp}

\end{document}